\documentclass[12pt,preprint]{aastex}

\long\def\symbolfootnote[#1]#2{\begingroup%
\def\thefootnote{\fnsymbol{footnote}}\footnote[#1]{#2}\endgroup}

\usepackage{ulem}
\usepackage{graphicx}

\shortauthors{Lagos et al.}

\begin{document}

\title{Star Cluster Complexes and the Host Galaxy in Three H\,{\sc ii} Galaxies: Mrk 36,
UM 408, and UM 461}

\author{P. Lagos\altaffilmark{1,2}}
\email{plagos@astro.up.pt}

\author{E. Telles\altaffilmark{2}}
\email{etelles@on.br}

\author{A. Nigoche-Netro\altaffilmark{3}}
\email{nigoche@iaa.es}

\and

\author{E. R. Carrasco\altaffilmark{4}}
\email{rcarrasco@gemini.edu}

\altaffiltext{1}{Centro de Astrof\'isica da Universidade do Porto, Rua das Estrelas, 4150-762 Porto, Portugal}
\altaffiltext{2}{Observat\'{o}rio Nacional, Rua Jos\'{e} Cristino, 77, Rio de Janeiro, 20921-400, Brazil}
\altaffiltext{3}{Instituto de Astrof\'isica de Andaluc\'ia (IAA), Glorieta de la Astronom\'ia s/n, 18008, Granada, Spain}
\altaffiltext{4}{Gemini Observatory/AURA, Southern Operations Center, Casilla 603, La Serena, Chile}

\begin{abstract}

We present a stellar population study of three H\,{\sc ii} galaxies 
(Mrk 36, UM 408, and UM 461) based on the analysis 
of new ground-based high resolution near-infrared J, H and K$_{p}$ broad-band and Br$\gamma$ 
narrow-band images obtained with Gemini/NIRI. 
We identify and determine relative ages and masses
of the elementary star clusters and/or star cluster complexes of the starburst regions
in each of these galaxies by comparing the colors with evolutionary
synthesis models that include the contribution of stellar continuum, 
nebular continuum and emission lines. 
We found that the current star cluster formation efficiency 
in our sample of low luminosity H\,{\sc ii} 
galaxies is $\sim$10\%. Therefore, most of the recent star formation
is not in massive clusters. Our findings seem to indicate that 
the star formation mode in our sample of galaxies is
clumpy, and that these complexes are formed by a few massive star clusters 
with masses $\gtrsim$10$^{4}$M$_{\odot}$. 
The age distribution of these star cluster complexes shows that 
the current burst started recently and likely simultaneously over short 
time scales in their host galaxies, triggered by some internal mechanism. 
Finally, the fraction of the total cluster mass
with respect to the low surface brightness (or host galaxy) mass, considering
our complete range in ages, is less than 1\%. 

\end{abstract}


\section{Introduction}

H\,{\sc ii} galaxies are dwarf galaxies undergoing an intense episode of star
formation (SF) that dominates their total optical luminosity. Most H\,{\sc ii}
galaxies overlap in their observed properties, with Blue Compact Dwarf (BCD) galaxies;
indeed, many are found in both samples. Typically, they have relatively
small physical sizes (with a few kpc in size), and the starburst regions spatially cover
the visible extent of the galaxy, making it difficult to access the presence of
an underlying (or host galaxy) stellar population. Originally, their low heavy
element abundance and the non-detection of an old population have given rise to the
question of whether they may be presently forming their first generation of
stars \citep[][]{S70}. 
Recent works, however, have shown that most
H\,{\sc ii} galaxies seem to present an underlying population from previous
episodes of SF \citep[e.g.,][]{Telles95,P96,T97a,C03}. 
So, the integrated light observed in H\,{\sc ii} galaxies is formed by 
the contribution of two main components, the young stellar population with ages of a few Myr to tens Myr
and an underlying population of intermediate (with ages of hundreds Myr) 
to old stars (with ages $\geq$ 1 Gyr).

The structure of the Interstellar Medium (ISM) found in H\,{\sc ii} galaxies
\citep[][and references therein]{L07} have profound implications for the origin
of the present starburst and the dominant large scale mode of SF. While many
high luminosity H\,{\sc ii} galaxies show evidence of morphological disturbances
that may be associated with interactions or ongoing minor mergers \citep[e.g.,][]{T97b,BO02}, 
less luminous and compact H\,{\sc ii} galaxies defy our attempt to find morphological signatures
of an external triggering agent \citep{T97b}.  In fact, H\,{\sc ii} galaxies seem to be typically isolated
and not associated with giant galaxies (Telles \& Terlevich 1995), and
their clustering properties seem to be similar to those of normal galaxies \citep{TM00}.
There have been, however, many attempts to search for HI companions \citep{T97}
and intrinsically faint optical companions \citep{N01,P01}. 
In these studies, it was concluded that galaxy interactions with both massive and dwarfs  
are probably the main mechanism that triggers SF bursts in BCD progenitors. 
Although interactions are often invoked to explain burst of SF in H\,{\sc ii}/BCD galaxies, 
it is possible that internal processes (e.g., gravitational cloud collapse 
and/or infall in conjunction with small perturbations) 
have been responsible for triggering the present episode of SF in at
least a significant fraction of dwarf galaxies \citep[see][and references
therein]{HE04,P04,T09,S11}.

It has been apparent for over a decade now, with the advent of the HST,
that the starburst regions in these galaxies are composed of a myriad of star
clusters \citep[e.g.,][]{B02}, with masses
$\gtrsim$10$^{4-5}$ M$_{\odot}$ and sizes of a few pc, that are typically more massive than normal
clusters in our Galaxy \citep{M95,CV94}.
These Super Star Cluster (SSCs) or
Young Massive Clusters (YMCs) were found originally in classical starburst
galaxies and in galaxies with evident signs of interaction or merger such as 
in the Antennae NGC 4038/4039 \citep{W99}, dwarf irregular galaxies such as NGC
1569 and NGC 1705 \citep{O94}, and other star-forming dwarf galaxies or BCD
galaxies such as M82 \citep{M05} and SBS 0335-052 \citep[][]{Th97,P98}. 
The formation of these massive clusters and the cluster mass function
(CMF) is directly connected to the SF processes in the galaxies, 
in the sense that different physical factors, in these galaxies, put constrains 
on the mass of the star clusters. The CMF appears to be a single power law (with index $\sim$-2), 
which implies the same SF mechanism for the massive clusters and their lower mass analogues.
This type of young and massive clusters were possibly formed
in high pressure conditions \citep[e.g.,][]{E97,B02}, hence the extremely
high pressure regions give rise to more massive and compact clusters. 
We know that SF 
in clusters is a common phenomenon in starburst galaxies, and that the massive clusters 
play an important role on the evolution of the ISM of their host galaxies, 
producing large scale structures such as supershells (or bubbles) and creating 
galactic winds that cause, in some cases, the blowout of freshly  produced 
metals from the galaxy into the intergalactic medium (IGM).
The similarity in mass and size between SSCs and Milky Way globular clusters
\citep[GCs; M$\sim 2\times10^{5}$M$_{\odot}$,][]{H91} suggests the possibility of
an evolutionary connection in the sense that SSCs are proto-globular cluster
systems. The classic example of a YMC is R136 at the center of  the 30 Doradus nebula
in the Large Magellanic Cloud. This cluster has a diameter of $\sim$1.6 pc, an
age of 1--2 Myr \citep{MH98} and an estimated mass of $\sim$2$\times$10$^{4}$
M$_{\odot}$ \citep[][and references therein]{W02}.

Apart from evidence for massive star clusters and its effect on the
surrounding medium, H\,{\sc ii} galaxies show an underlying low surface
brightness (LSB) component likely to represent their host galaxies, whose stellar
population is a product of previous episodes of SF. This underlying
stellar host generally extends a few kpc from the central star-forming 
regions, showing regular and elliptical isophotes with red colors indicative
of an evolved stellar population. The host galaxy in H\,{\sc ii} galaxies has
been studied in the literature based on the analysis of both optical \citep[e.g.,][]{T97a} and
near infrared (near-IR) images \citep[e.g.,][]{C03,N03,N05} and spectroscopy \citep[e.g.,][]{R00,W04}. 
\citet{T83} observed, in the near-IR, a sample of BCD galaxies and showed that the  
light in these galaxies is due to the presence of K- and M-giant stars. 
The study of the structural parameters of the host galaxy 
(exponential scale length ($\alpha_{0}$) and central surface brightness ($\mu_{0}$)) 
has been commonly used in the literature to evaluate the
evolutionary relations between the different types of dwarf galaxies
\citep[e.g.,][and references therein]{P96,T97b,G05}. The derivation of the ages and
spatial distribution of these stellar populations is the first step towards
establishing the evolutionary state and the SF history of these
galaxies.

We present near-IR broad-band J, H and K$_{p}$ and Br$\gamma$
narrow-band images of three low luminosity H\,{\sc ii} galaxies: Mrk 36, UM 408, and UM 461. 
Our aim is to describe the properties of the star clusters or complexes which are
distinguishable with our superb ground-based high spatial resolution images with NIRI on
the Gemini North telescope as well to determine the structural properties of the underlying 
galaxy using surface photometry. These observations in combination with a proper assessment of
recent stellar population synthesis models allow us to put some constraints on
the recent  and  past history, and  the dominant large scale mode of SF in these
galaxies.

The paper is arranged as follows: 
In \S~\ref{data} we describe the sample, the observations, and data reduction.
In \S~\ref{results} we present the results and derive the properties of the star clusters/complexes 
and the LSB component. In \S~\ref{discussion} 
we discuss our results and finally in \S~\ref{conclusions} we summarize our conclusions.
 
\section{Sample, observations and data reduction}\label{data}

\subsection{Our sample}\label{sample}

We targeted two galaxies Mrk 36 and UM 461, that are particularly rich in
giant star-forming knots and one compact H\,{\sc ii} galaxy, UM 408, 
with less evidence of multiple knots of SF. 
The light collecting power of the Gemini North telescope and the high spatial resolution 
permitted by NIRI are an excellent combination for resolving the star cluster or/and 
star cluster complex populations in these compact galaxies. In the following 
paragraphs we describe the main properties of our sample of galaxies.

\textit{Mrk 36 (Haro 4, UGCA 225)} is a compact H\,{\sc ii} galaxy showing at
least two large star-forming knots \citep{L07} in the south-eastern part of the
galaxy. \citet{T83} reports integrated near-IR colors  (de-reddened) of 
$J-H=0.38$ and $H-K=0.42$, while \citet{HE06} reports integrated colors of
$J-H=0.422 \pm 0.100$ and $H-K=0.543 \pm 0.310$.
The regions of largest H$\beta$ line emission
\citep{L07} and HI maps \citep{BA04} coincide, indicating that the current SF is
restricted to the dense region of the parental cloud. Bravo-Alfaro et al. argue
that a transient encounter between Mrk 36 and the neighboring spiral Haro 26
could explain both the SF in the former and the pronounced warp in
the latter. Finally, recent radio observations with the VLA in 1.4, 4.9 and 8.4 GHz by \citet{RG07}
have shown that Mrk 36 has a nearly flat radio spectrum dominated by thermal
emission, similar to the regions detected by \citet{JK03} in Henize 2-10. This
may be an indication that the first contribution of the synchrotron emission to
the low frequency emission, due to the first supernova (SN) explosions, has not yet
appeared. These integrated properties show definite evidence of a very young starburst. 

\textit{UM 408} is a compact galaxy with a projected size of $\sim$1kpc.
Although this galaxy was resolved as a single H\,{\sc ii} region in previous
studies \citep[e.g.,][]{G03}, two giant regions in the central part of the galaxy
were detected by \citet{L09},
with ages of $\sim$5 Myr and stellar masses of $\sim$10$^4$M$_{\odot}$. Using
GMOS-IFU observations \citet{L09} showed that the metal content in the ISM of
this galaxy is well mixed and homogeneously distributed throughout
the galaxy, in the same way as in other dwarf galaxies. 

\textit{UM 461 (PGC 037102)} is a well studied H\,{\sc ii} galaxy. This object has
a very compact and bright off-center nucleus, some small
regions spread along the galaxy \citep{N03}, and an external envelope that is
strongly distorted towards the south-west. \citet{D01} and \citet{N03} have
obtained integrated near-IR colors that differ significantly from each other,
with J-H=0.99 and H-K=-0.68 and J-H=0.47 mag and H-K=0.2 mag, respectively.
\citet{TE95} proposed that UM 461 was formed together with UM 462 by tidal interaction. But the HI
maps of UM 461 and UM 462 do not show that these galaxies are tidally
interacting, therefore it is unlikely that these objects induced the star
formation on each other \citep{V98}.

Finally, the presence of He \,{\sc ii} $\lambda$4686 emission bump of Wolf--Rayet (WR)
stars has been reported in the 
integrated spectra of Mrk 36 and UM 461 \citep{C91}.

\subsection{Observations, data reduction, and calibration}\label{obser_data_cal}

Broad-band J(1.25$\mu m$), H(1.65$\mu m$), K$_{p}$(2.12$\mu m$), and narrow-band
Br$\gamma$(2.17$\mu m$) images were obtained using the NIRI instrument on the 
Gemini North telescope on 2005 August 02 (UM 408), November 24 (Mrk 36), and December 29 (UM 461). 
We used the f/6 camera which provides a field of view of $\sim$120\arcsec$\times$120\arcsec using the 
1024$\times$1024 pixels ALLADIN InSb detector, with a pixel scale of 0\arcsec.116 on side.
The observations were performed under photometric conditions. 
Table \ref{observation_log} lists all observational parameters and values adopted in
this work. In this Table, Column (1) gives the object name. Columns (2) and (3) give
the $\alpha$ and $\delta$ coordinates (J2000), respectively. Column (4) gives the
observed heliocentric velocity (vel.) and the 3K CMB (Cosmic Microwave
Background radiation) corrected distance from NED. Columns (5) and (6) give the
oxygen abundance and extinction c(H$\beta$) for these galaxies obtained from the literature. 
Column (7) gives the galactic extinction E(B-V) from the extinction maps of \citet{S98}. 
In Column (8) we present the date of observation. Column (9) shows the filter used in each
observation and Columns (10) and (11) give the exposure time in seconds,
considering the different coadd exposures and the mean air mass of each 
observation, respectively. Finally, Column (12) shows the 
instrumental constants C$_{\lambda}$ of the transformation equation for flux calibration
as described below in this section.

The data were reduced following the standard procedures for near-IR 
imaging using the Gemini/NIRI package version 1.8 inside IRAF\footnote{IRAF: the Image Reduction and 
Analysis Facility is distributed by the National Optical Astronomy Observatories, which is operated by the 
Association of Universities for Research in Astronomy, In. (AURA) under cooperative agreement 
with the National Science Foundation (NSF).}. For each filter, a normalized flat was constructed from
the flat images observed with the calibration unit with the shutter closed (lamps off) and with the 
shutter open (lamps on). The bad pixel mask was constructed by identifying the bad pixels in the flat
images with shutter off. The sky images were constructed from the raw science images by identifying
all objects in each frame, masked out, and averaging the remaining good pixels (images were observed 
with a dither offsets of 10\arcsec~and 20\arcsec~ for all galaxies). The raw science images were
processed by subtraction the sky  on a frame-by-frame basis and dividing by the normalized flat field 
images. The final flat-fielded, sky subtracted images were then registered to a common pixel position
and median combined.

In order to obtain a photometric calibration, we
observed two standard stars: FS130 (GSPC P264-F) and
FS21 (GD140) for the observing run of Mrk 36, FS4 (SA93-317) and FS6 (Feige 22)
for UM 408, and FS19 (G162-66, LTT 3870) and FS20 (G163-50, LTT 4099) for the
observations of the galaxy UM 461. For each filter, the photometric calibration was obtained using the
following equations:

\begin{equation}
 m_{\lambda_{0}}=m_{\lambda_{i}}+\chi K_{\lambda}+C_{\lambda},
\end{equation}

\begin{equation}
 m_{\lambda_{i}}=-2.5\times log(c/t),
\end{equation}

\noindent where m$_{\lambda_{0}}$ is the magnitude in the standard system,
$m_{\lambda_{i}}$ is the instrumental magnitude with c the number of counts and
$t$ the exposure time, $\chi$ the air mass, K$_{\lambda}$ the extinction
coefficient and C$_{\lambda}$ the instrumental constant ($\lambda$=J, H,
K$_{p}$). The instrumental constant C$_{\lambda}$ was calculated as the average
value from the standard stars. Given the limited
number of standard stars observed, we have used the average value of extinction
to the ``Mauna Kea''  observatory (MKO) $K_{\lambda}$=0.1, 0.06 and 0.09
\citep{K87} to the filters J, H and K, respectively. The instrumental 
C$_{\lambda}$ constants used for each galaxy are listed in the last column of Table \ref{observation_log}.
To compare the observed K$_{p}$ measures with evolutionary
synthesis models, we transformed the K$_{p}$ into K magnitudes using the
relation K$_{p}$= K + 0.22$\times$(H-K) \citep{WC92}.
We selected star clusters/complexes by visual inspection of the images 
and using \textit{daofind} in DAOPHOT. We fitted the detection threshold properly,
with a value of 2.5$\sigma$ above the local background, in order to detect sources in the starburst region that 
in the case of H\,{\sc ii}/BCD galaxies spatially cover
the visible extent of the galaxies. All the clusters or complexes in this
study have been detected in the J, H and K$_{p}$ band. The final catalog of objects
is shown in  \S~\ref{analysis_SSC}.

The Br$\gamma$ images were calibrated using a procedure similar to that
described in \citet{L07}. We convolved the stellar Spectral Energy
Distribution (SED) with the response of the narrow-band filter. To do this we used the SED,
appropriately scaled, of stars with the same spectral type as our standard stars
obtained from the literature. Finally, we subtracted the continuum,
estimating their contribution from the modeled SED for a given age of the regions obtained below
in \S~\ref{model}.  

Table \ref{mag_int} shows the observed integrated magnitudes for all the galaxies, 
uncorrected for Galactic extinction. 
The magnitudes were calculated by measuring the flux inside  fixed apertures of radii 14\arcsec, 
6\arcsec, and 12$\arcsec$  for the galaxies 
Mrk 36, UM 408, and UM 461, respectively. Foreground objects have been masked out.

\subsection{Completeness limits}\label{completeness}

\begin{figure}[ht]
\centering
\includegraphics[scale=.30]{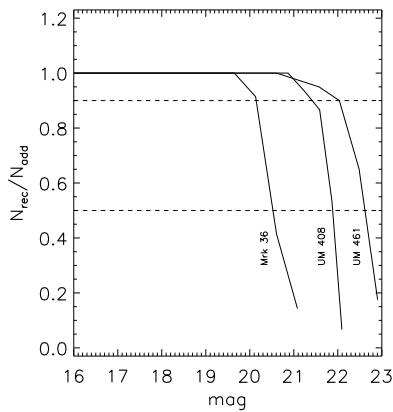}
\caption{Completeness profiles (N$_{rec}$/N$_{add}$) in the K$_{p}$ filter for star clusters in each of our studied galaxies,
from the left to the right, Mrk 36, UM 408 and UM 461, respectively.
The dashed lines show the recovered fraction of 90\% and 50\%.
\label{comple}}
\end{figure}

We quantified our detection limits and test the reliability of the derived magnitudes 
using a series of completeness tests by adding artificial extended objects to our images.
First, we created an empirical point spread function (PSF) for isolated point-like sources 
in the NIRI images. Given that the observed field of view (FOV) is relatively 
small ($\sim$120\arcsec$\times$120\arcsec), 
there is a possibility that these point-like sources are non-stellar.
So these objects were selected by eye and we ruled out sources with irregular PSF shapes.
Using this information we added 20--50 artificial extended objects distributed in a regular square array. 
Models of these extended sources were generated using mksynth in BAOLAB \citep{La99} and 
were added to the science images.
Magnitudes were randomly assigned to each position from 16.00 mag to 23.00 mag with an interval of 0.5 mag. 
Finally, we compared the number of added and recovered sources in each galaxy. 
The estimated completeness fraction for objects 
as a function of magnitude for the K$_p$ band is shown in Figure \ref{comple}.
The completeness goes down to 20.2 mag in Mrk 36,
21.5 mag in UM 408 and 22.03 mag in UM 461 with $\sim$90 \% of the objects recovered in the K$_{p}$ band.

\section{Results}\label{results}

\begin{figure*}[!ht]
\centering
\includegraphics[scale=.40]{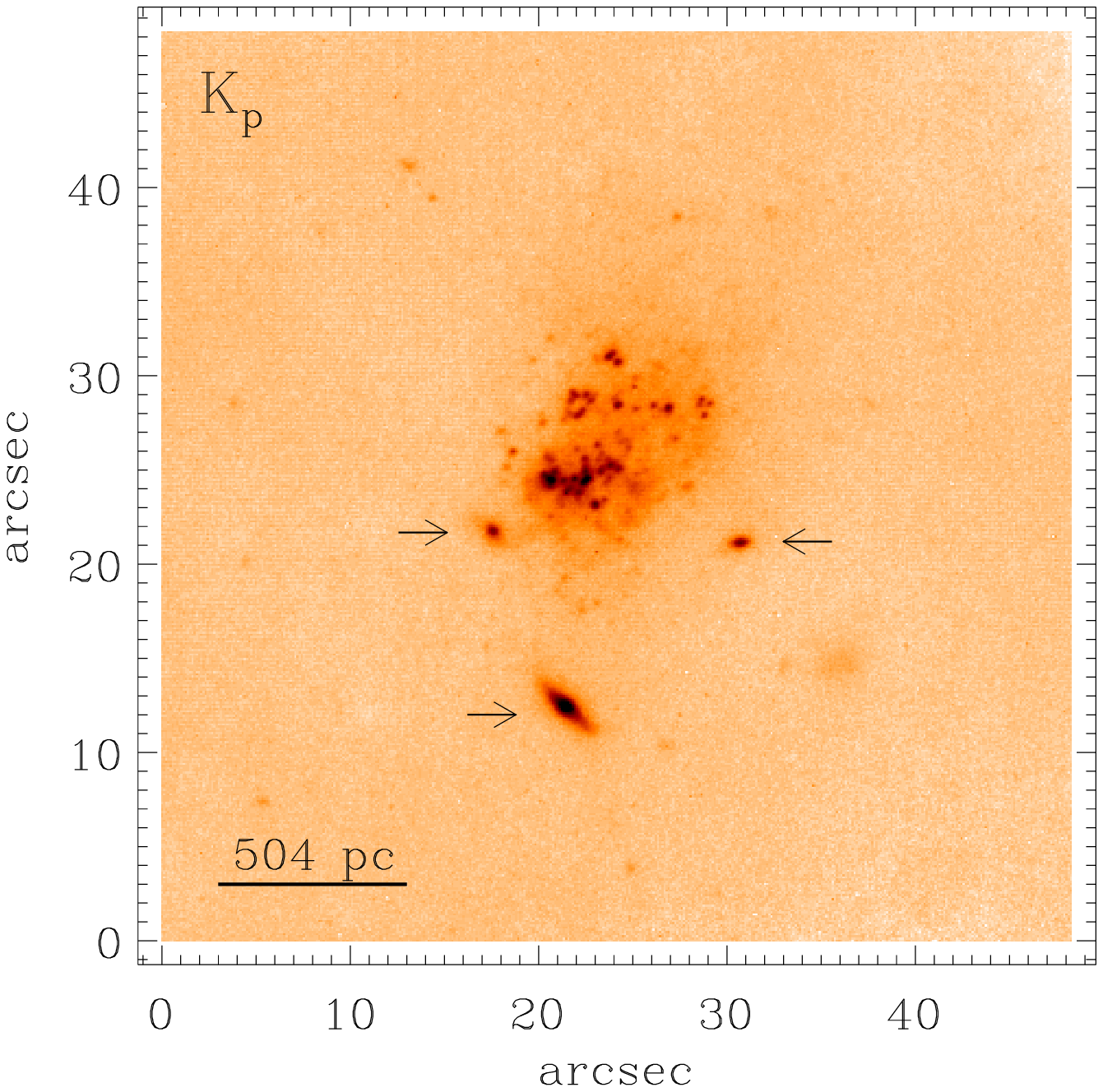}
\includegraphics[scale=.40]{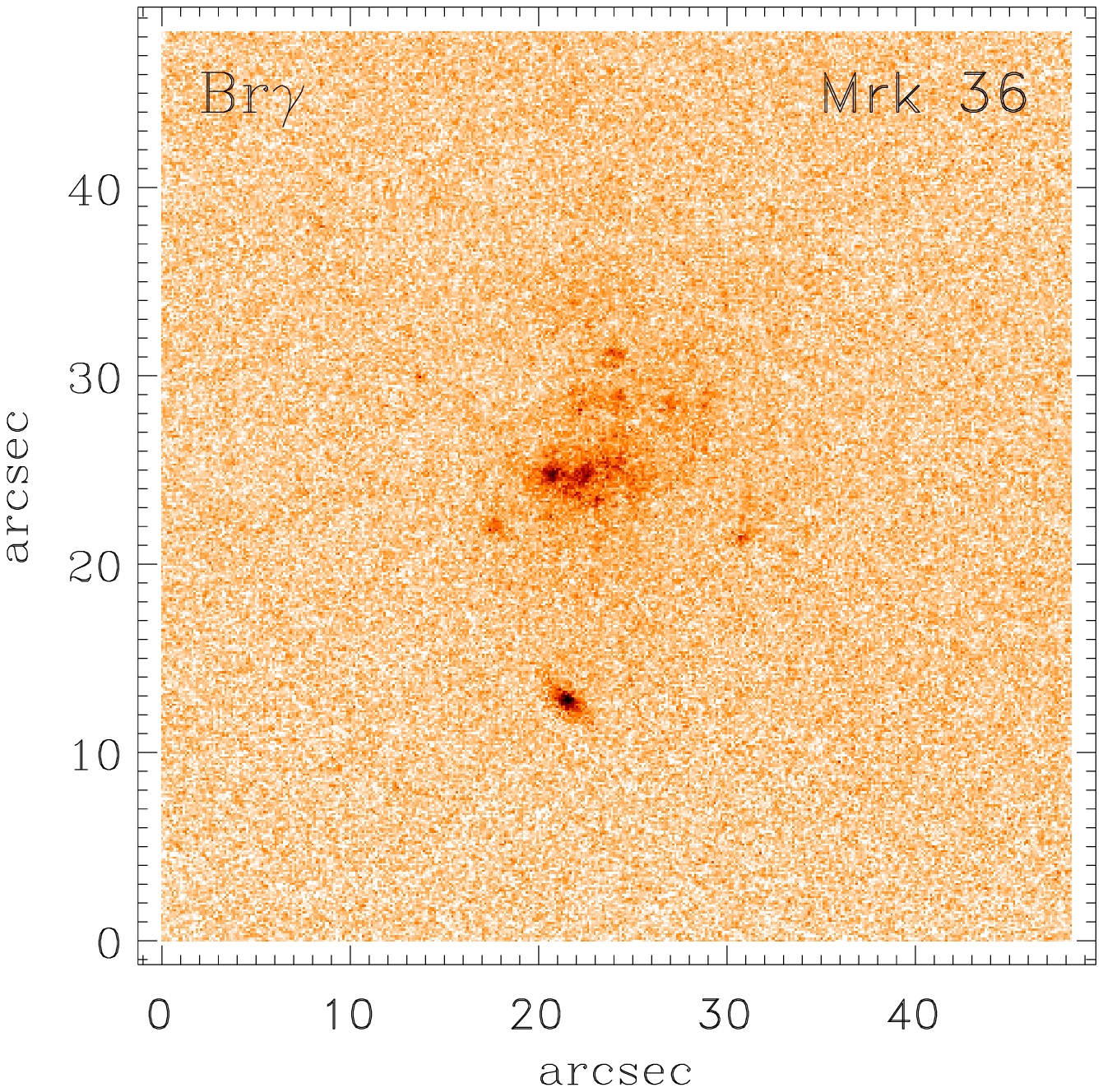}\\
\includegraphics[scale=.40]{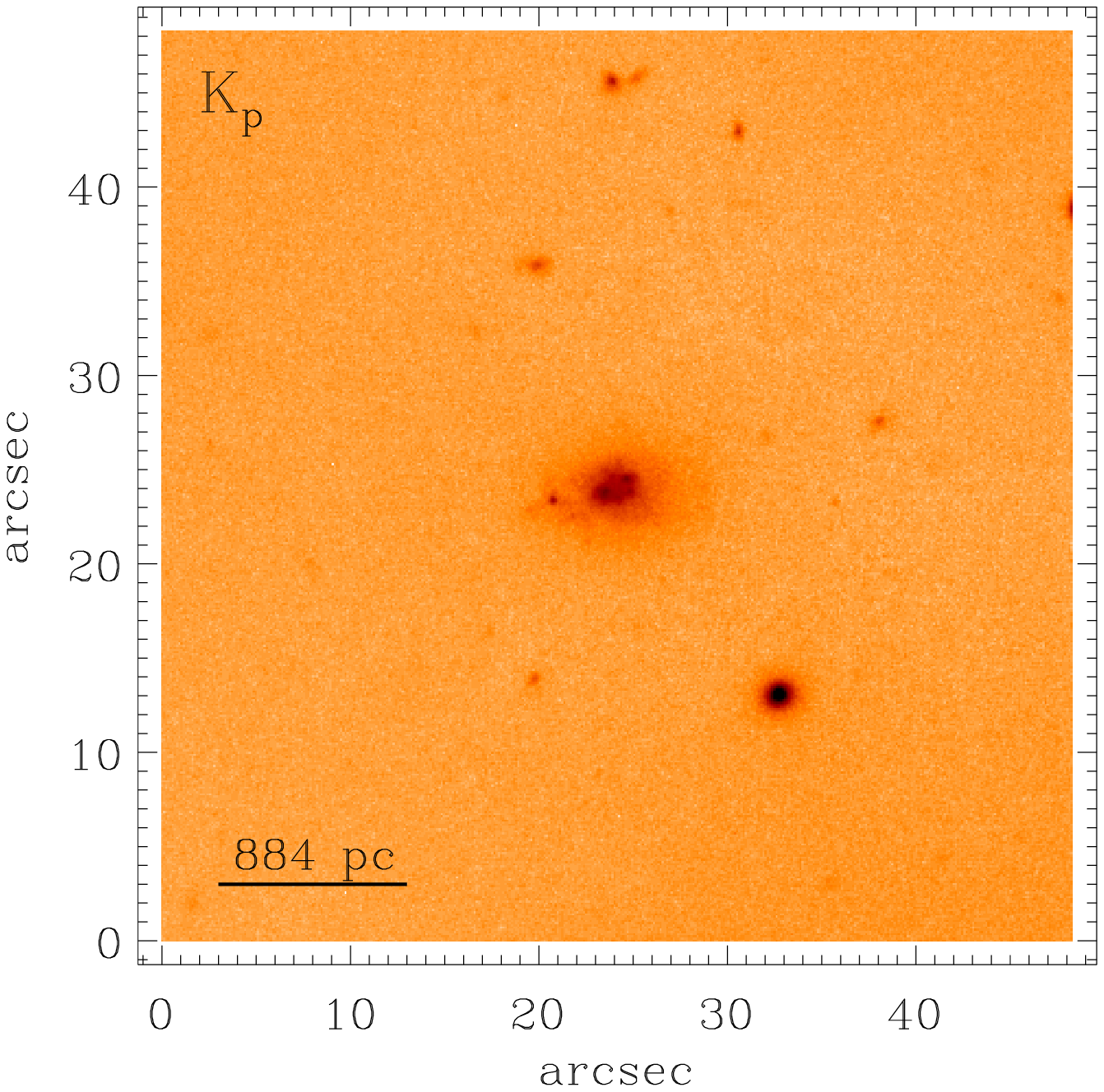}
\includegraphics[scale=.40]{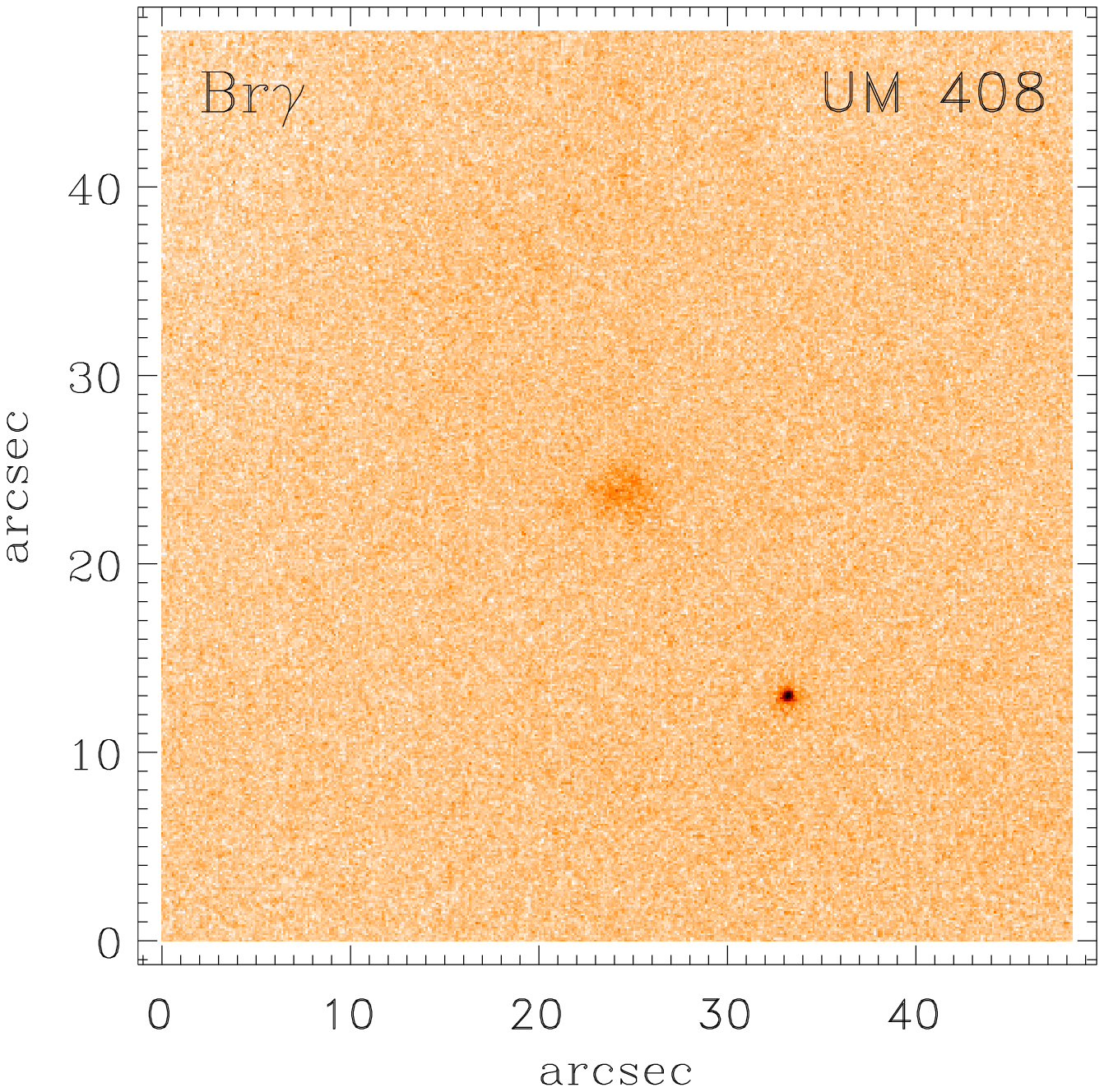}\\
\includegraphics[scale=.40]{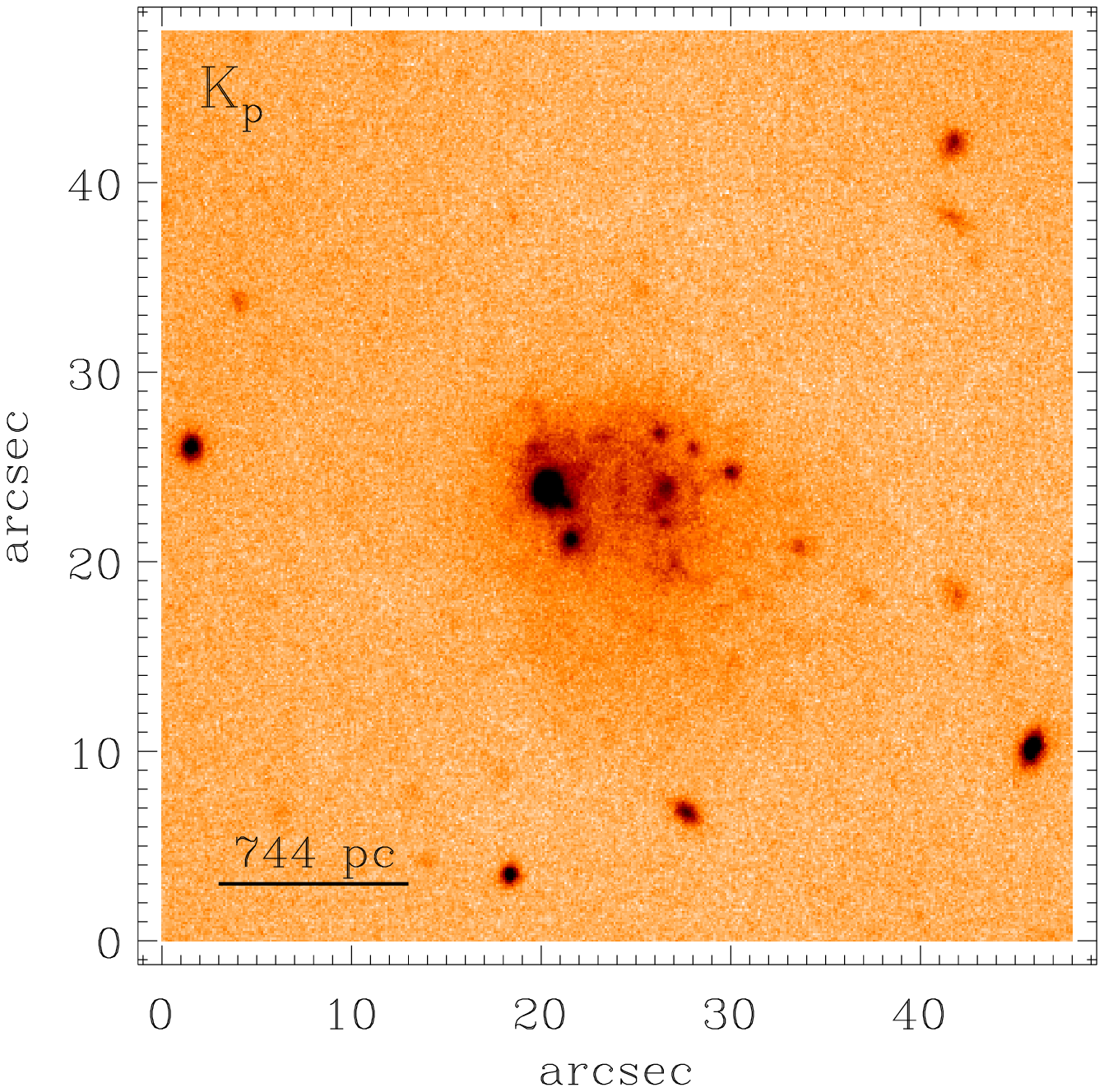}
\includegraphics[scale=.40]{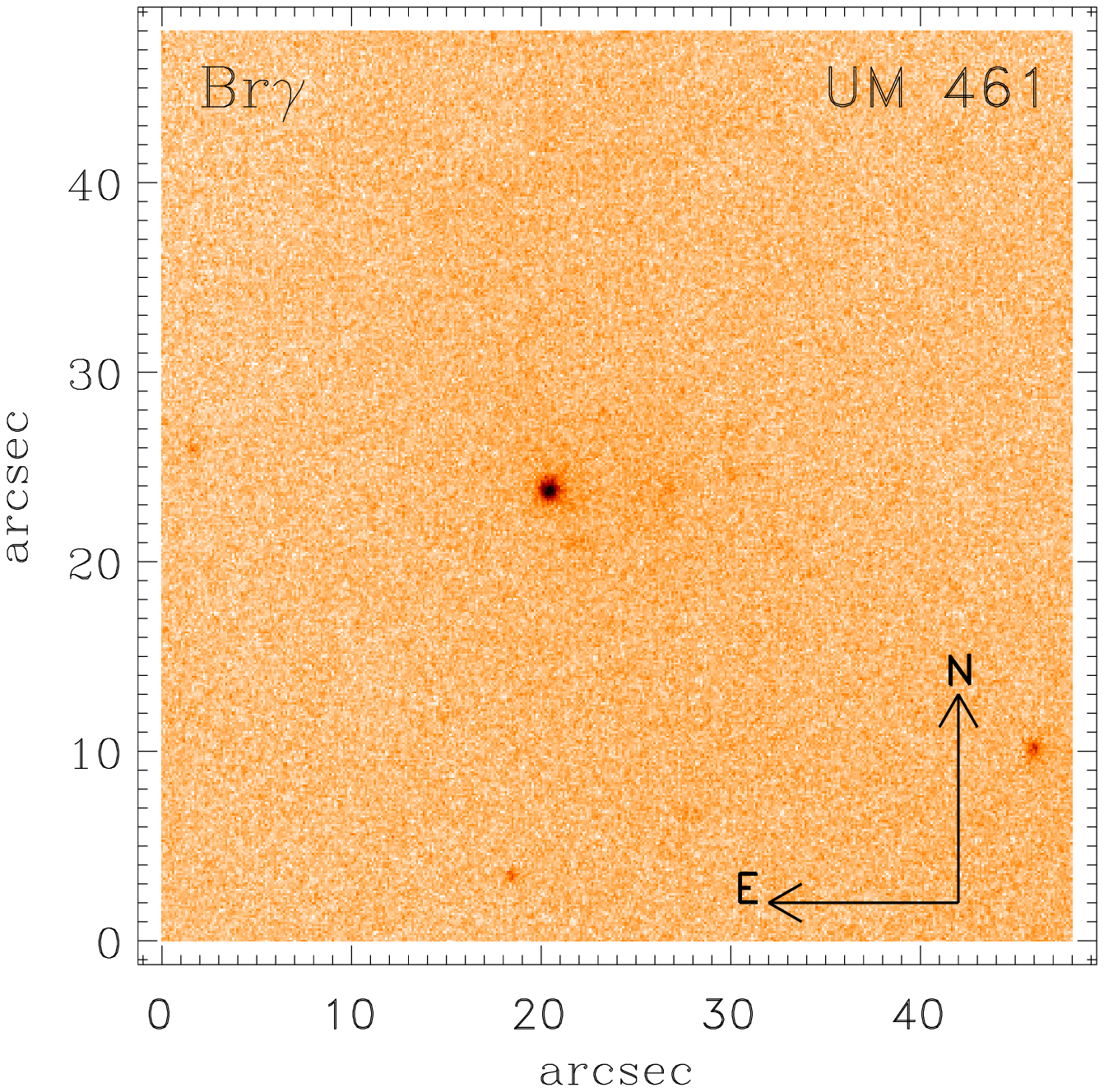}
\caption{K$_{p}$ and Br$\gamma$ images (not continuum subtracted) of the galaxies Mrk 36 (top), UM 408 (middle) and UM 461 (bottom)
with NIRI at Gemini North. The field of view of each image is 48$\arcsec\times$48$\arcsec$. 
The images are displayed on a logarithmic scale. The arrows in the
K$_{p}$ image of Mrk 36 point to three background galaxies close to
the galaxy. North is at the top and east to the left.\label{observations}}
\end{figure*}

Figure \ref{observations} shows the three galaxies observed and studied in this
work (Mrk 36, UM 408, and UM 461) in the K$_{p}$ and Br$\gamma$ bands,
respectively. All galaxy images reveal the presence of bright regions and/or star cluster complexes,
given the apparent sizes of these regions in Figure \ref{observations},
surrounded by LSB envelopes. Br$\gamma$ emission is detected in all galaxies,
but only the brightest regions show clear evidence of intense emission given by our
detection limit. The unprecedented high spatial resolution 
images obtained for Mrk 36 allow us to identify, for the first time using ground-based
telescopes, the elementary structures within the starburst. Mrk 36 shows a plethora of
small clusters, distributed over the entire extension of the galaxy, but with a 
previously unresolved concentration in the central knot. The Br$\gamma$
emission shows a peak in this region, indicating the young nature of these
clusters. We also observed the presence of other groups of bright clusters in the
K$_{p}$ image located in the northern region of the galaxy with weaker Br$\gamma$ emission.
Here we named these two groups of regions as complex I and complex II, respectively.
The galaxy UM 408 appears quite compact and regular, and the
K$_{p}$-band image shows that the star-forming regions are distributed in a clearly defined
ring-like geometry, composed of five almost regularly spaced regions
where the current SF is occurring. The largest region is
located in the eastern part of the galaxy. We observe the presence of another region outside the
central part of the galaxy. The Br$\gamma$ emission in this galaxy is very regular and weak. 
Finally, the K$_{p}$-band morphology of UM 461 is similar to that
reported in previous works \citep[e.g.,][]{N03}. The brightest region is
off-center and located in the eastern part of the galaxy.
This region is extremely compact and bright in Br$\gamma$.
We detected a number of regions spread throughout the body of the galaxy.

\subsection{Properties of the star clusters/complexes }\label{analysis_SSC}

\subsubsection{Near-IR colors}\label{nearIR_colors}

\begin{figure}[!ht]
\centering
\includegraphics[scale=0.5]{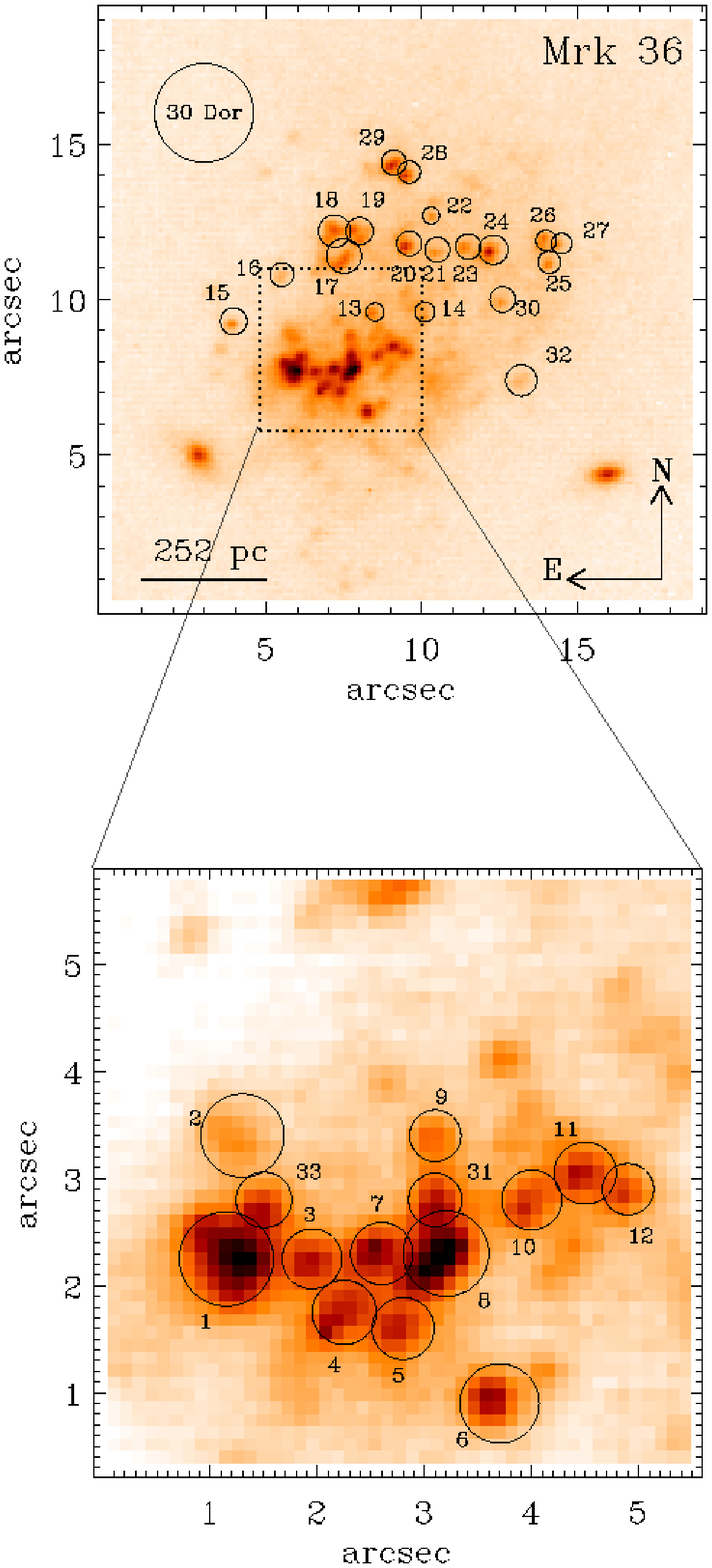}
\includegraphics[scale=0.5]{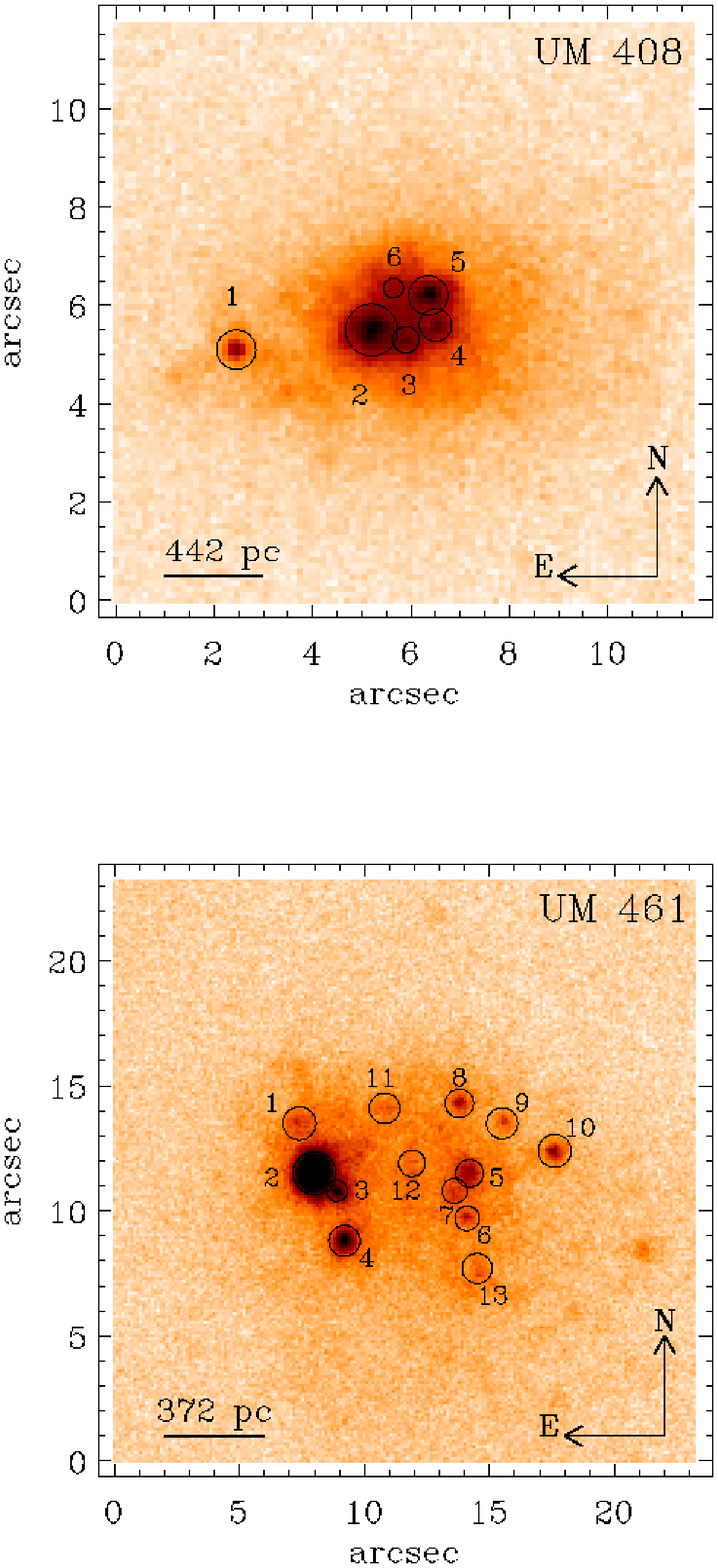}
\caption{Images of Mrk 36, UM 408, and UM 461 in the K$_{p}$ filter. Circles mark the position (and
the apertures used for photometry) of each star cluster/complex identified in the galaxies. 
The clusters have been labeled with the designation used throughout the paper (see Table \ref{clusters_data}). 
In the upper panel of Mrk 36 we show the position of each cluster identified in the northern region of the galaxy. 
The circle at the corner of this panel represents the size of the nebular region of
30 Doradus \citep{W91}. The lower panel shows the clusters in the central
region of Mrk 36 (dashed square in the upper panel). \label{clusters}}
\end{figure}

For the color analysis we considered the objects obtained from our
catalog in \S~\ref{data} that are brighter than m$_{Kp}\sim$21 mag. 
We detected 33 regions in Mrk 36, 6 in UM 408 and 13 in UM 461, respectively. 
All regions identified in the K$_{p}$-band images of the galaxies are shown in Figure
\ref{clusters}. Given the seeing of $\sim$0\farcs4--0\farcs5 we expect these regions to
have diameters less than 25 pc in Mrk 36, 88 pc in UM 408, and 37 pc in UM 461. 
So these diameters are upper limits, and the elementary cluster population 
in our samples of galaxies are barely resolved. 
Many of the detected regions are not individual star clusters but rather blends of several 
individual star clusters with sizes similar to that in other star cluster complexes
in the literature \cite[e.g.,][]{M09}. 
In the case of Mrk 36 the detected regions are compatible with the sizes
of some young star clusters detected in other H\,{\sc ii}/BCD galaxies \citep[e.g., M82;][]{M05}, 
so these regions can be considered as individual star clusters, although likely some of these sources are blended. 
In any case, young clusters are not formed in isolation but rather are found in cluster complexes \citep{Z01,La04}.
Hence, the properties derived are the luminosity-weighted mean value of the complex.    
We have to bear in mind this caveat in our analysis but we are still able to derive 
the light weighted properties of these knots of SF and evaluate the clumpiness properties 
of the large scale mode of SF in these galaxies. In Figure \ref{clusters} we also show the size
\citep[$\sim$200 pc;][]{W91} of the whole nebular region of 30 Doradus to compare with the sizes
of the clusters in Mrk 36.

\begin{figure}[!ht]
\centering
\includegraphics[scale=0.5]{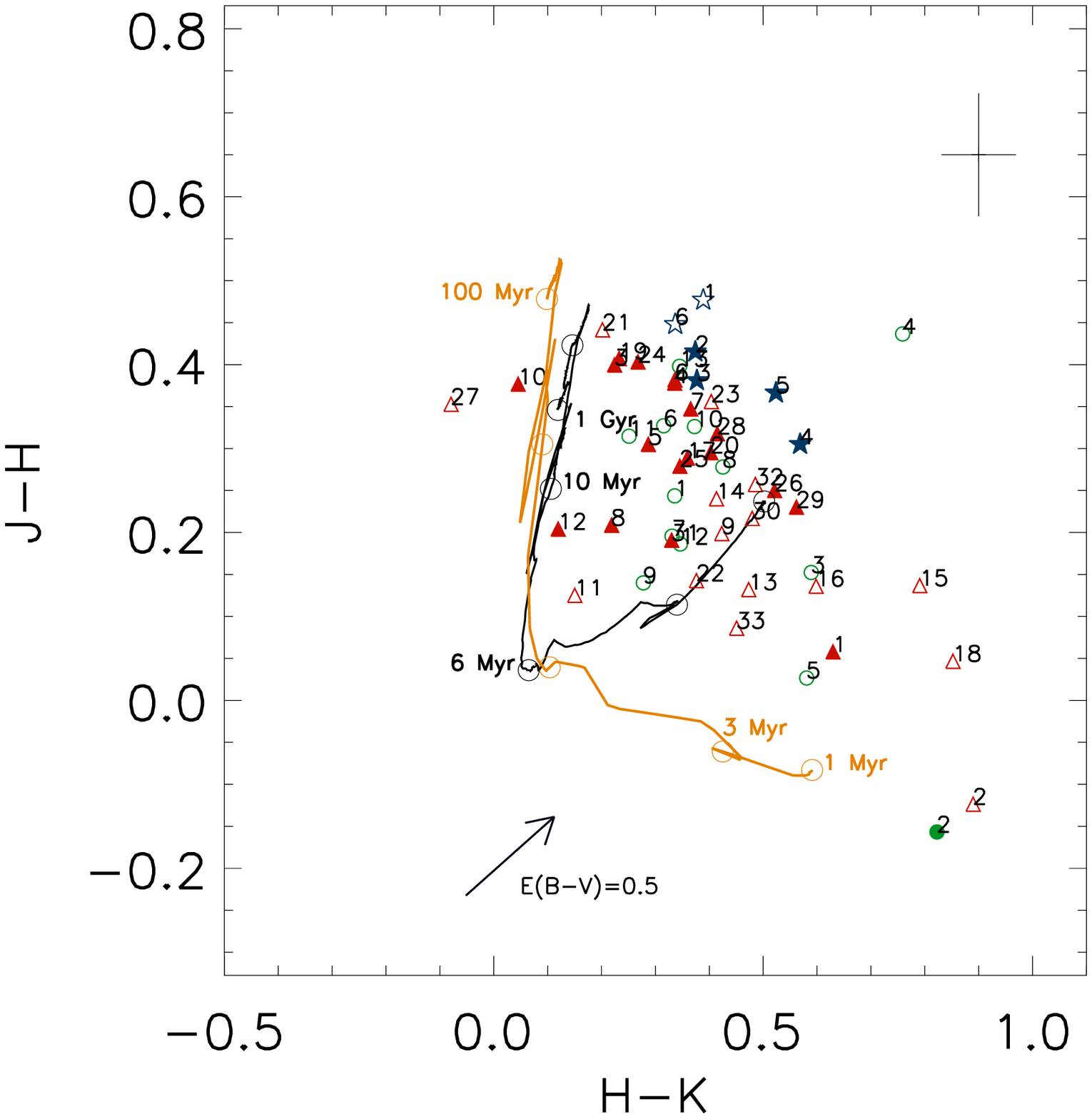}
\caption{Observed near-IR color--color diagram J-H vs. H-K for the star clusters/complexes in our
sample not corrected for extinction. The lines show the evolutionary tracks of
these colors from STARBURST99 (black line; this model includes stellar and
nebular continuum) and GALEV models (orange line; this model includes stellar continuum,
nebular continuum and the contribution of nebular emission lines) for a metallicity
Z=0.004. Red triangles represent the observed values for
clusters in Mrk 36, the blue stars represent the values for UM 408 and 
green circles represent the values for clusters/complexes in UM461. Filled symbols
indicate the detection of Br$\gamma$ in the regions. The error bars show
the average error value for the colors. Open circles along the tracks indicate ages
of 1, 3, 6, 10, and 100 Myr. An additional circle indicating an age of 1 Gyr is
included in the model of STARBURST99.\label{clusters-colors}}
\end{figure}

We measured the flux of the individual star cluster/complexes
in all filters (including Br$\gamma$) using circular apertures with the program APER in IDL (an adapted 
version of the task DAOPHOT in IRAF). For each aperture, we considered only pixels with
3$\sigma$ flux above the background. 
For each cluster we calculated the colors
J-H and H-K (after the transformation of K$_{p}$ into K magnitudes). Figure
\ref{clusters-colors} shows the color--color diagram (J-H vs. H-K) for all star clusters/complexes.
In this figure we illustrate the evolutionary tracks of these colors using
STARBURST99 \citep[black line;][]{L99} and GALEV \citep[orange
line;][]{K09} single stellar population (SSP) models, for a Kroupa IMF, and metallicity Z=0.004 (more details in
\S~\ref{clusters:age}). The triangles represent the observed values for
clusters in Mrk 36, the stars represent the values for the star cluster complexes in UM 408, and circles
represent the values for the star cluster complexes in UM 461. Filled symbols
indicate the detection of Br$\gamma$ in these regions.
Finally, we compare our observed apparent magnitudes with the ones found in the literature. 
The only one cluster/complex found in the literature is the region \#2 in UM 461 where \citet{N03}
obtained m$_{J}$=17.30 mag, that agree within the errors with our value of m$_{J}$=17.34 mag. 

Table \ref{clusters_data} lists the photometric values for each star cluster/complex identified 
in our sample of galaxies (see Figure \ref{clusters}). 
In this Table, Column (2) shows the identification number.
Column (3) shows the aperture considered to obtain the photometry in arcsec.
Columns (4), (5), and (6) give the observed photometry in the J, H, and
K$_{p}$ bands for each cluster, respectively. The $^{\dagger}$ symbol over the
identification number indicates that Br$\gamma$ emission was measured. Finally,
Columns (7), (8), and (9) give the extinction E(B-V), age in units of Myr, and the
stellar mass in units of 10$^{4}$M$_{\odot}$ of each star cluster or complex derived
from the best fit comparison of the observed colors with chosen evolutionary
synthesis models as described below in \S~\ref{clusters:age}.

\subsubsection{Obtaining the physical properties of the star cluster/complex population}\label{clusters:age}

\subsubsubsection{Models}\label{model}

Absolute age dating of young star clusters in starburst
galaxies has proven to be a very challenging task, due to the additional emission by dust plus gas, as well
as the effects of extinction and metallicity. In relative terms the determination of
ages and masses of the star cluster population within these galaxies can show some insight
on the overall dominant mode of recent SF. In order to show these effects
we derived the extinction and ages using our color--color diagrams (see Figure~\ref{clusters-colors}) 
by comparing the near-IR colors for each star cluster or star cluster complex with two independent SSP models:  
STARBURTS99 models (model I; this model includes pure stellar and nebular continuum) and 
GALEV (model II; this model includes stellar continuum, nebular continuum, and emission lines from warm ionized gas) 
for an instantaneous burst of SF. So we derived the ages by calculating
the best fit of the models (using the chi-square method) to the observed colors for the model, 
varying the extinction E(B-V) from 0 to 1 in steps of 0.05 mag.
We used the Galactic extinction curve given by \citet{C89} assuming R$_{V}$=3.1. 
The star cluster/complex masses were derived using the absolute magnitudes M$_{k}$ of each cluster 
as compared with the models. In the following paragraphs we describe the main
properties of the models used in this work.

\textit{Model I}: We considered the STARBURST99 model, which includes pure stellar and nebular continuum, 
for metallicity Z=0.004 ($\sim$1/5Z$_{\odot}$), and Geneva evolutionary stellar tracks, assuming a
Kroupa IMF ($\propto$M$^{-\alpha}$) with $\alpha$=1.3 for stellar masses between
0.1 to 0.5M$_{\odot}$ and $\alpha$=2.3 for masses between 0.5 and 100M$_{\odot}$
for a total mass of 10$^{6}$M$_{\odot}$. More details about the physics of the model in \citet{L99}.

\textit{Model II}: We compared  our data with the GALEV  evolutionary track for metallicity Z=0.004 and 
a Kroupa IMF (0.1-100M$_{\odot}$) including stellar continuum, nebular continuum, and gas emission. 
The GALEV models used in this work were kindly
provided to us by Ralf Kotulla. These models were run using the Geneva
evolutionary tracks with a minimum age and time resolution of 0.1 Myr, unlike
the models available on the GALEV website which use the Padova isochrones and
have a minimum age and time step of 4 Myr. The models also use a fraction of
visible mass, which is used for the standard models available on the Web page, and which are
thus twice as bright as the Padova models for the same mass. This does
not affect any of the colors, but we must keep this in mind when comparing
masses. The flux of the hydrogen lines were computed using atomic physics 
and the production rate of ionizing photons, whereas non-hydrogen line strengths 
are computed using metallicity-dependent line ratios relative to H$\beta$.
More details about the input physics is given in \citet{K09}. 

\subsubsubsection{Extinction, masses and cluster/complex ages}

\begin{figure}[!ht]
\epsscale{1.0}
\plotone{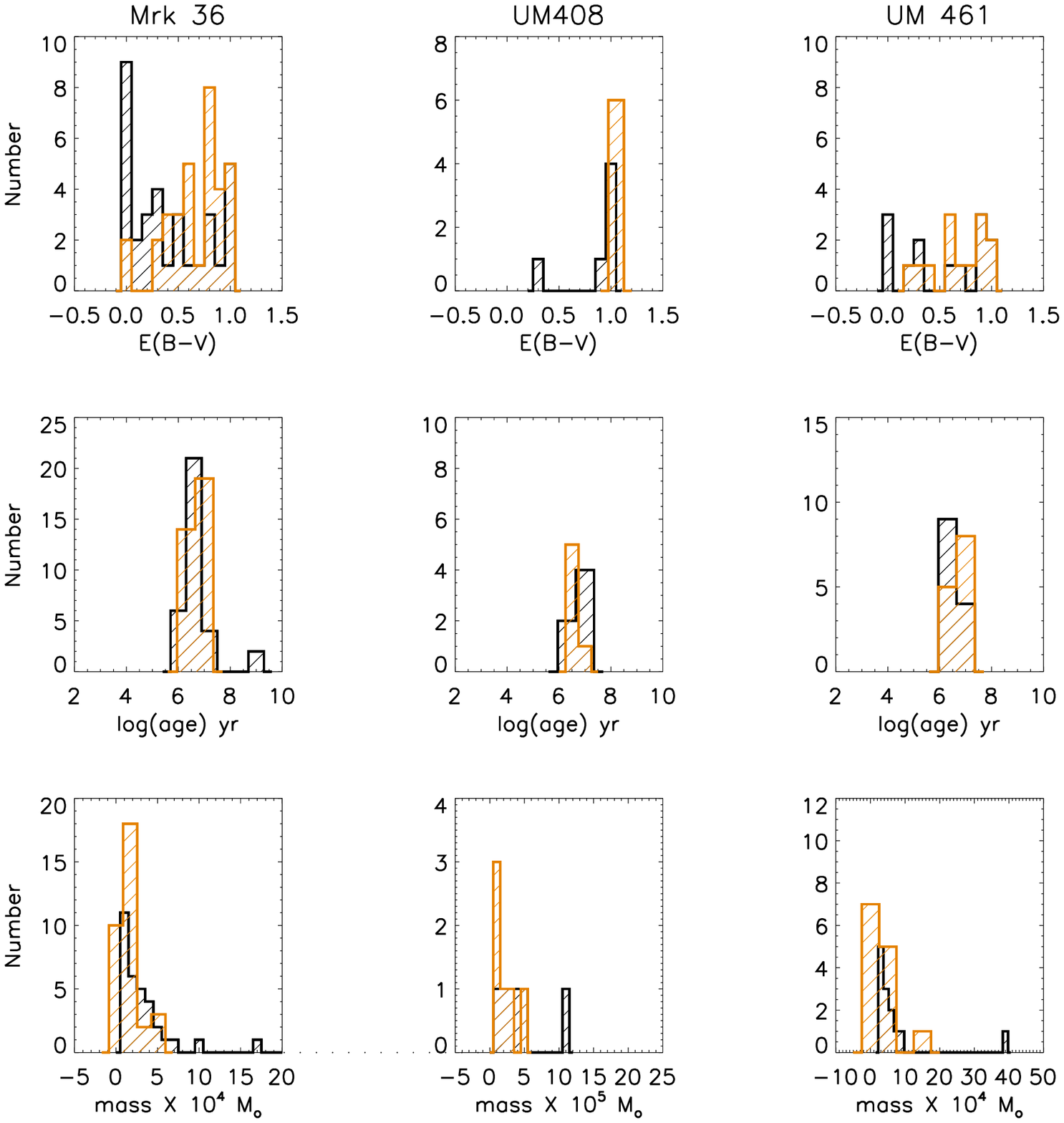}
\caption{Extinction, age and mass distribution of our sample of star clusters/complexes,
obtained using the STARBURST99 (model I; black distribution) and 
GALEV (model II; orange distribution) models. For more details see
\S~\ref{clusters:age}. \label{cluster_properties}}
\end{figure}

The two models (model I and II) displayed on the J-H  versus H-K color--color diagram (Figure \ref{clusters-colors}) show a
different path for ages $\lesssim$6 Myr. The track for model I and the extinction vector, in the corner of this figure, 
are perpendicular to the model II. Therefore, estimates of extinction could be significantly different when
we use the two cases. 
In Columns (7), (8), and (9) of Table \ref{clusters_data} we show the
E(B-V), age and mass of the star clusters/complexes, using the two different models
described above. In Figure \ref{cluster_properties} we show the distribution of
E(B-V), ages, and masses for the star clusters/complexes detected in this work. In this Figure, the
black distribution corresponds to the results obtained using STARBURST99 and the orange distribution
to the those obtained using GALEV. From this figure it is clear that the extinction distribution 
is shifted to higher values when we use the model that include the contribution of nebular emission lines, 
meanwhile, the ages estimated using this model (model II) appear to be slightly 
younger and less massive than those obtained using model I.

We can ask now, what is the most appropriate model to represent 
the ages and masses of our sample of star clusters/complexes? 
The SED of the observed star clusters/complexes in some H\,{\sc ii}/BCD
galaxies show a clear excess in the near-IR, 
that is reflected in red H-K colors with respect to the models.
This observed red excess had been reported earlier by \citet{V00,V02,H03}, \citet{J04}, and more
recently by \citet{R08} and \citet{A10,A11} in the star cluster population of other H\,{\sc ii}/BCD
galaxies. In fact \citet{A10} showed that the I and H bands are significantly affected by a 
similar excess. \citet{R10} and \citet{A10} showed that this red excess clearly introduces
a systematic offset in all of the derived parameters when all filters, from UV-optical to near-IR
are included in the determination of the observed SEDs. This suggests that
the models commonly used in the literature to calculate the properties of the clusters,
models that include pure stellar continuum, are inadequate for age-dating studies \citep{P98,A03,R10}.    

Some of the causes analyzed in the literature that have been proposed to explain this
excess are: i)  there may be an important contribution from nebular continuum
and line emission \citep[e.g.,][]{P98,V02,R10}, ii) hot dust emission \citep[e.g.,][]{V00,R08}, iii) the presence of young Red
Super Giant (RSG) stars in  older clusters, not properly modeled at low
metalicity \citep[][and references therein]{MM01,V07}, and iv) other sources as for example extended red emission 
\citep[ERE;][]{WV04}, produced by photoluminescence process and the presence 
of a population of young stellar objects \citep[YSOs;][]{A11} still surrounded by circumstellar
disks.
 
Since there is no degeneracy in the color space between age and extinction at young ages (Figure \ref{clusters-colors}), 
at least qualitatively, model II explains the colors of young star clusters \citep{R10} given that only this
model is able to reproduce the trend seen in the data, in Figure \ref{clusters-colors}, at young ages. 

The presence of nebular continuum and emission lines in the near-IR can have a large
impact on the inferred properties (affecting the mass determination) of the star clusters in H\,{\sc ii}/BCD
galaxies, and models that include this effect are the most appropriate in the
study of young stellar population with ages $\lesssim$6 Myr.
In fact, near-IR spectra of Tol 35, Tol 3, and UM 462 presented by
\citet{V02} clearly show recombination lines of HeI and HI (Br$\gamma$, Br$\delta$ and Pa$\beta$),
and excited lines. The line emission detected by Vanzi et al.
is sufficient to produce their observed broad-band excess.
Although hot dust and the presence of RSGs may be important factors that can contribute to the near-IR excess, 
we assume that the excess in the SF regions in our sample of H\,{\sc ii} galaxies
is mainly produced by nebular continuum and emission lines, hence we can use our near-IR photometric bands and model II, that 
include this contribution, in order to estimate the properties of the detected star clusters/complexes.

\subsubsection{Uncertainties in the age determination}

\begin{figure}[!ht]
\centering
\includegraphics[scale=.3]{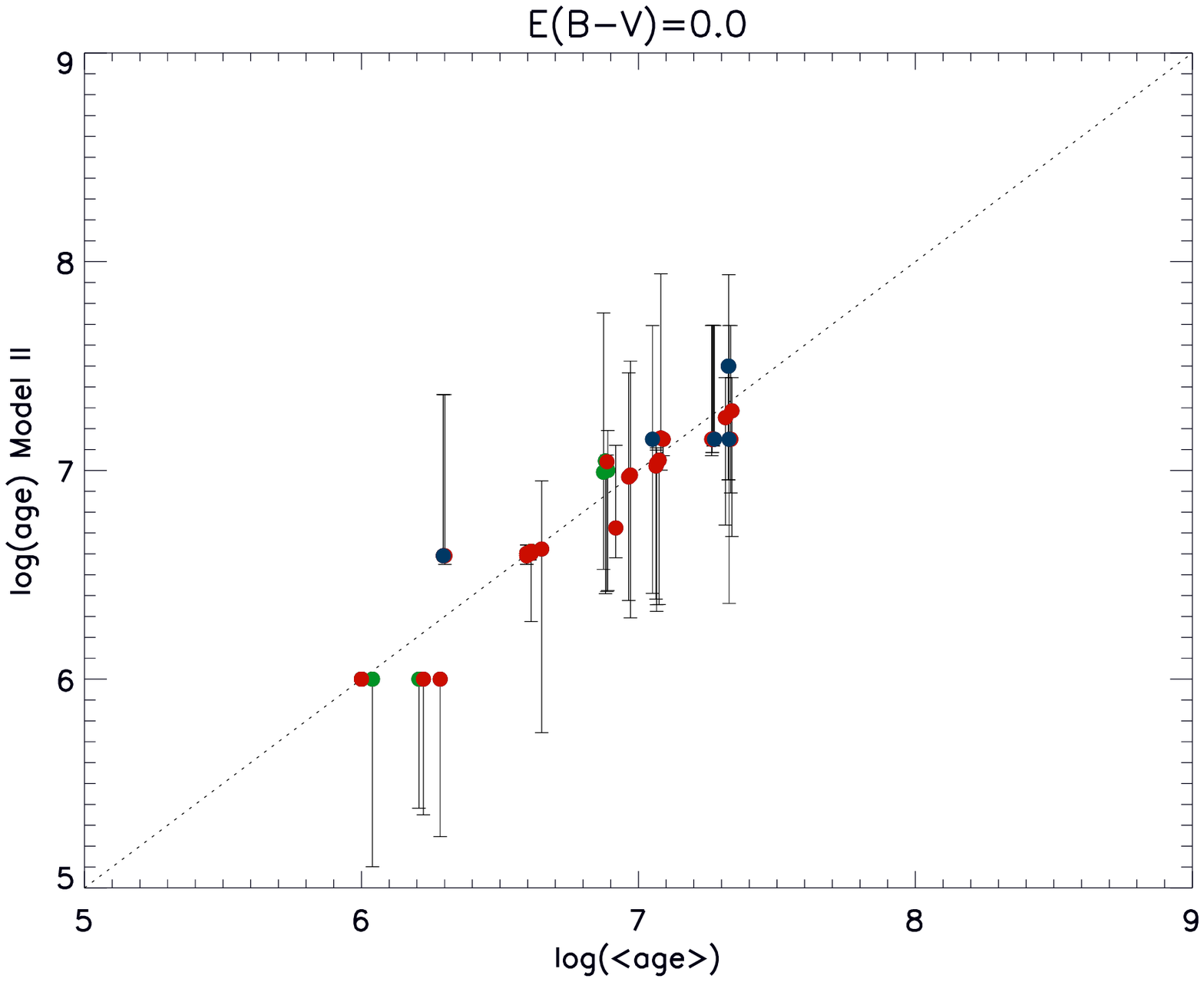}
\includegraphics[scale=.3]{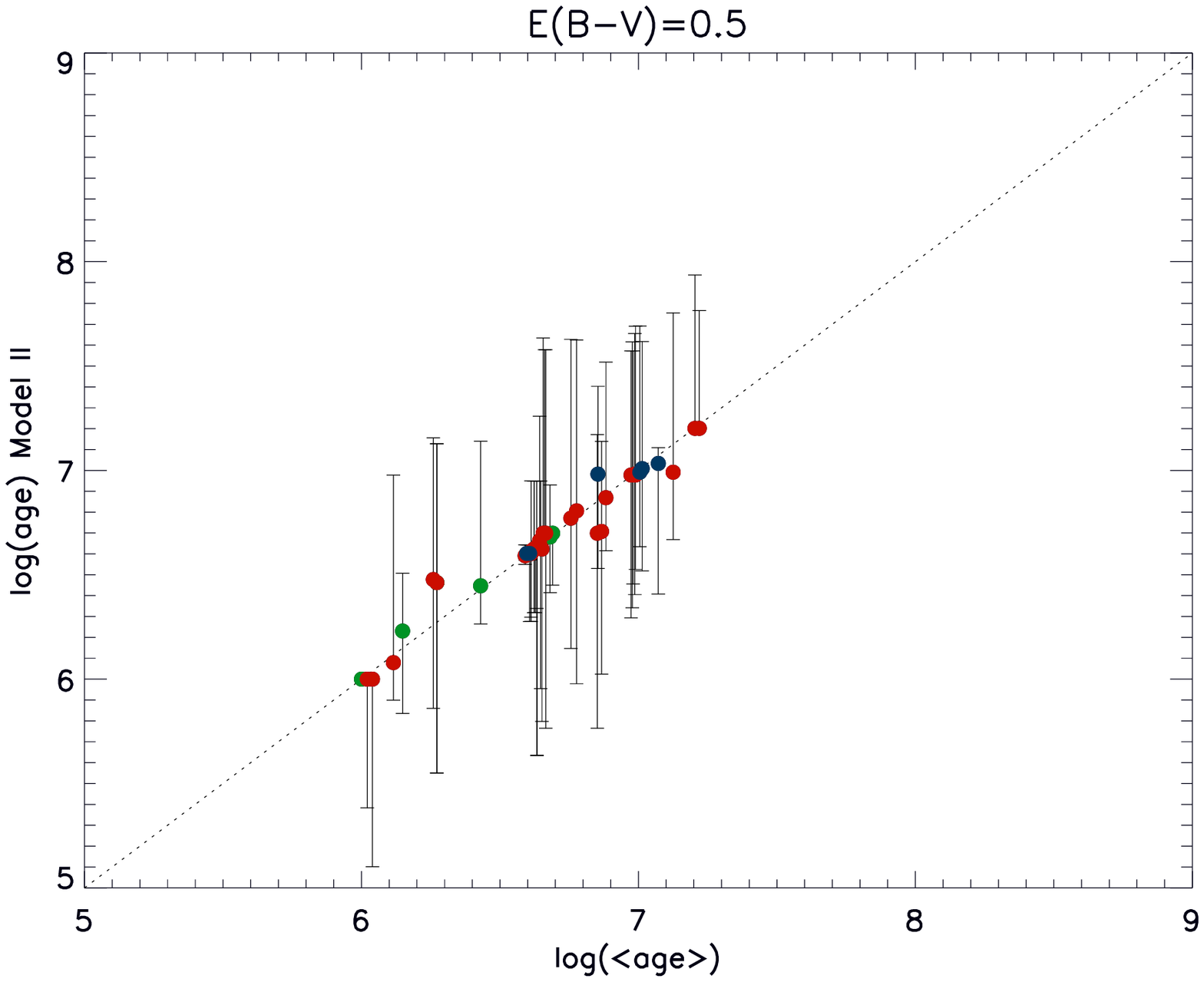}
\includegraphics[scale=.3]{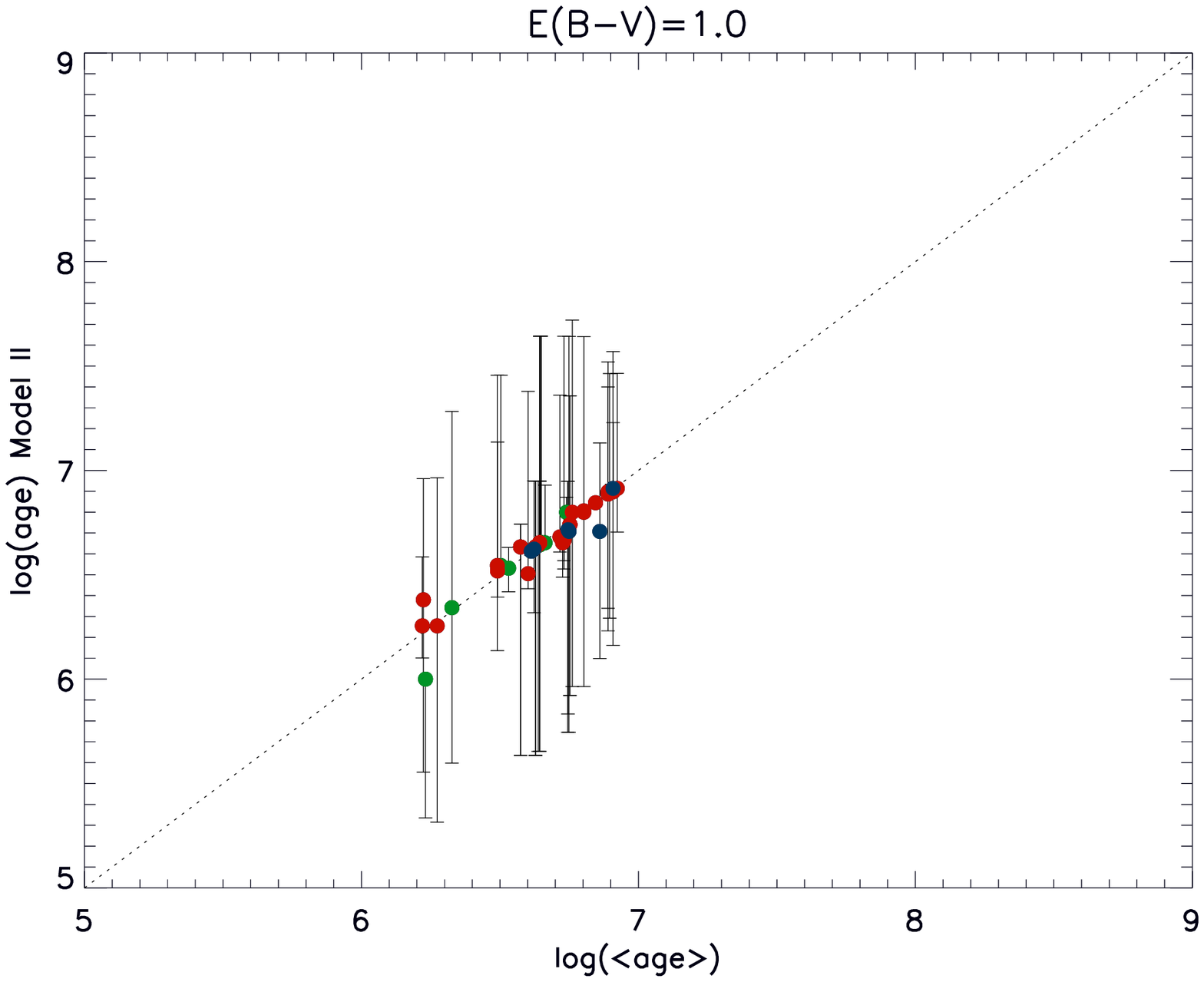}
\caption{Best fit ages (listed in table \ref{data}) from model II for Mrk 36 (red), 
UM 408 (blue) and UM 461 (green) for different extinctions E(B-V) as a function of the mean age 
obtained varying the colors by their typical observational errors. The errors in
this figure were calculated from the difference between the ages from the observed colors and the ones obtained by
displacing these colors by their 1$\sigma$ uncertainties. \label{test}}
\end{figure}

In order to estimate the uncertainties in our age calculation procedure, we calculate the best fit ages while varying 
the colors by their typical observational errors (as listed in Table \ref{data}).
In Figure \ref{test} we show the ages obtained from model II for three different extinctions E(B-V)=0.0, 0.5 and 1.0
as a function of the mean age ($<$age$>$) obtained varying the colors by their errors. 
We showed that in the worst case (if all the errors conspire), 
they can modify the solutions, but not changing significantly the properties of the sample.  
In this figure, we can see that for high extinction the clusters/complexes are better fitted than for lower
extinctions, producing that the best fits are obtained at higher values of reddening.
This is an indication that the starburst regions or complexes are dominated by highly extinguished and 
very young star clusters.

A major concern in our method stems from the fact that we are using near-IR bands 
to obtain the properties of the star cluster/complexes.
The determination of absolute ages/masses of star clusters and star cluster complexes 
using near-IR colors, alone, could be highly uncertain and highly model dependent.  
It seems that better constraints of the properties of the star clusters/complexes are obtained 
with a large broad band coverage plus emission line equivalent widths. 
However, we depend on the improvement of population synthesis models and the contraints of the input 
parameters and ingredients, as stellar tracks, libraries at different metallicities, and inclusion of the
effects of nebular emission, dust, etc.
Given these uncertainties, we cannot obtain absolute ages, but our analysis, based on these observations 
give us a qualitative view of a rather homogeneous star cluster population. 
Relative ages and masses can, however, be useful in order to indicate
the mode of SF at galactic scales. 

We note that future high-resolution ground-based or space telescope observations using
UV-UBVRIJHK broad bands are needed in order to obtain the SED of each star cluster 
or/and complex in our sample of galaxies and then constrain absolute ages and quantify the 
near-IR excess produced by other mechanism, e.g., hot dust, ERE, YSOs, RSGs in metal poor ambients, etc.

\subsubsection{Summary of the obtained properties}

In summary, we found that the star cluster population in Mrk 36 is massive with 
estimated masses of $\sim$10$^{4-5}$M$_{\odot}$. We detected, in this galaxy, a few 
clusters with masses of $\sim$10$^3$M$_{\odot}$. 
Given our detection limits, lower mass clusters are likely not to be detectable.
Meanwhile, the star clusters complexes in UM 408 and UM 461 have masses from $\sim$10$^{4}$M$_{\odot}$ to
$\sim$10$^{6}$M$_{\odot}$. The age distribution shows that the detected star clusters/complexes are very 
young with ages less than $\sim$10 Myr in Mrk 36 and $\sim$5 Myr in UM 408 and UM 461, respectively. 
The two star clusters with colors consistent with ages $>$10-100 Myr, in Mrk 36,
are likely old star clusters as is, likely, the case of the cluster \#27. Other possibility,  
is that these objects are field galaxies not properly resolved.
As we mentioned previously, some of the star cluster complexes are highly extinguished and 
young. In Mrk 36, the less extinguished clusters, in most cases, are the youngest star clusters 
(mainly in complex I). This feature is less clear in UM 408 and UM 461, where our resulting properties 
are averaged over a wider region. In the case of UM 408, our results agree with the ones obtained by \cite{L09}, 
where using GMOS--IFU spectroscopy they found that the highest values of extinction in the c(H$\beta$) map
are displaced from the peaks of H$\alpha$ emission. The position of the peaks of extinction found by Lagos et al.
are correlated with the position of our detected star cluster complexes. 
This feature suggests that the current starburst episode is sweeping the gas and dust out of the center
into the surrounding regions.

Finally, from a comparison of our results with the ages and masses of other star clusters 
in H\,{\sc ii}/BCD galaxies studied in the literature, we see that the ranges 
are similar. 
For example, the analysis of the cluster population in UM 462 has
revealed ages between 4.7 and 10 Myr \citep{V03} with masses range from 1.2 to
7.2$\times$10$^{5}$M$_{\odot}$. In the case of Haro 11 \citet{A10} found that
30\% of the clusters have masses $>$10$^5$M$_{\odot}$, arguing that these clusters
qualify as SSCs. Using HST Imaging Spectrograph (STIS) long-slit far- and
near-ultraviolet spectra, \citet{C04}  studied a local sample of SSCs in WR
starburst galaxies, including Mrk 36. They estimated an age $<$1 Myr for the
SSC that coincides with the position of our cluster \#1, the most luminous cluster in Mrk 36.
Given that WR features has been reported for Mrk 36 and UM 461 in the literature and
since long slit spectra are taken centered on the
brightest regions, we expect the brightest clusters in Mrk 36 to be really 
young clusters with upper age limit in the range $\lesssim$2-5 Myr \citep{GV95}. 
If the detected clusters, in Mrk 36, are really young, and given that
their ages, masses and sizes (with diameters less than 25 pc) are similar to the
properties obtained in other star clusters, particularly in interacting and star
forming galaxies, we suppose that some of the most massive star clusters, found
in Mrk 36, can be considered candidates to be SSCs. The sizes 
of the regions in the other two H\,{\sc ii} galaxies studied here
are characteristic of star cluster complexes. High spatial resolution observations in space
are needed in order to resolve the elementary entities that
constitute the starburst population in UM 408 and UM 461.  

\subsubsection{The spatial distribution of the star cluster/complexes}\label{spatial}

\begin{figure}[!ht]
\epsscale{1.0}
\plotone{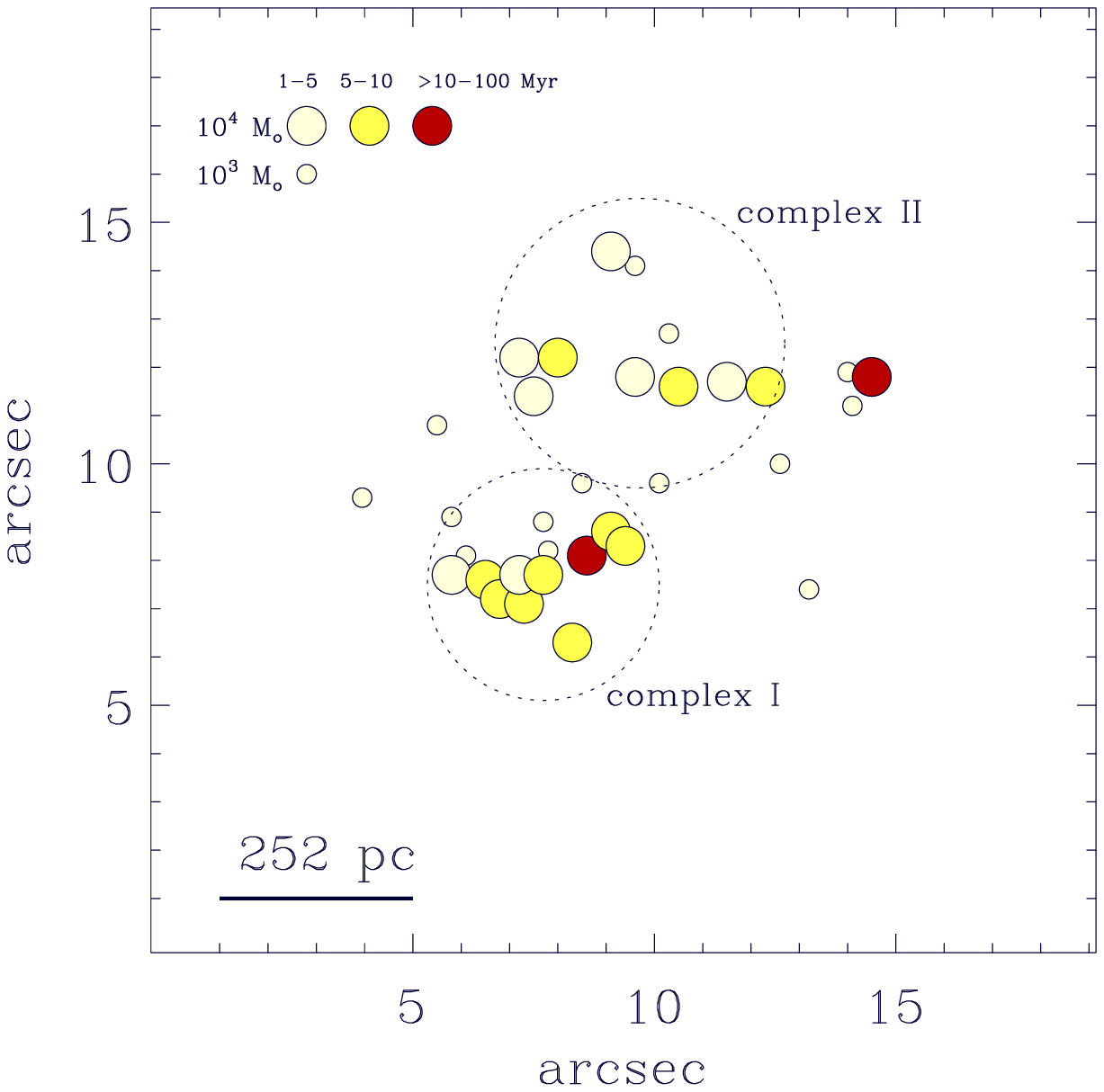}
\caption{Spatial distribution of the star clusters in Mrk 36. The different
circle sizes correspond to masses $\sim$10$^{3}$M$_{\odot}$ and $\sim$10$^{4}$M$_{\odot}$ considering
three ranges in age: 1-5, 5-10 and $>$10-100 Myr.}\label{DIST_AGE_MASS}
\end{figure}

In a self-propagating SF model \citep{G78} the gas expansion caused by stellar-wind and SNe
shock waves will triger the next generation of stars. 
If this scenario of SF is plausible, we must  derive an age trend of
the star clusters, or age sequence of some groups of star clusters with respect to the spatial position. 
To illustrate this, we show in Figure \ref{DIST_AGE_MASS} the position of the clusters in the galaxy Mrk 36.
The star clusters, in this figure, are divided into two groups 
with masses $\sim$10$^{3}$M$_{\odot}$ and $\sim$10$^{4}$M$_{\odot}$, considering
three ranges in age: 1-5, 5-10 and $>$10-100 Myr.
The spatial distribution of the star clusters/complexes, their relative ages and
spatial position in the host galaxy show that there is no clear age trend, meaning that
the young regions were not triggered by the action of the older ones.

In order to estimate the propagating timescale
of the feedback from the star-formation activities, we calculate the crossing time 
t$_{cross}$=R/v$_{prop}$, where R is the size of the system and v$_{prop}$
the propagation or expansion velocity. Typically, this velocity 
varies from $\sim$10 kms$^{-1}$ to $\sim$100 kms$^{-1}$ in dwarf galaxies \citep[e.g.,][]{Va07}.
Denoting by R the distance between the central or brightest cluster or complex to the 
other clusters, we found that the propagation of SF is possible within the complexes in Mrk 36,
for high expansion velocities, given that the t$_{cross}<\Delta$age (with a mean t$_{cross}$ and 
$\Delta$age$\sim$3 Myr in Mrk 36), but not on galactic scales because t$_{cross}>\Delta$age. 
For low velocities the difference between
the ages of the central region (in complex I) with respect to the more distant clusters (in complex II) 
are $\Delta$age$\ll$t$_{cross}$ (with a mean t$_{cross}\sim$30 Myr in Mrk 36),
suggesting that the propagation of SF between the complexes is improbable. 

Given that the star clusters/complexes are practically coeval, the star
formation was produced simultaneously within time scales of the order
of $\Delta$age on galactic scales.
More likely a global SF mechanism is responsible for the present
SF activity in galactic scales in some H\,{\sc ii} galaxies, 
whereas, self propagating SF on scales of $\lesssim$100 pc is still
possible within the individual complexes.

\subsection{Properties of the LSB component}\label{analysis_LSB}

\subsubsection{Surface Brightness Profiles and Structural parameters}

In order to obtain the surface brightness profile, for each galaxy,
we fitted ellipses to the isophotes using the IRAF task \textit{ellipse}. 
First, we manually masked the bright regions identified in the K$_{p}$-band images, except the central
ones, and replacing the values by the average of the adjacent regions. Then, we approximated the
initial ellipse centers, ellipticities, and position angles, allowing these
parameters to vary freely with radius during the fitting process. 
When the routine cannot proceed in the iterations as it reaches the lowest surface brightnesses,
the ellipse task stops, and we fix the center, ellipticity, and the position 
angle with these values to produce our light profiles. 

\begin{figure}[!ht]
\centering
\includegraphics[scale=0.35]{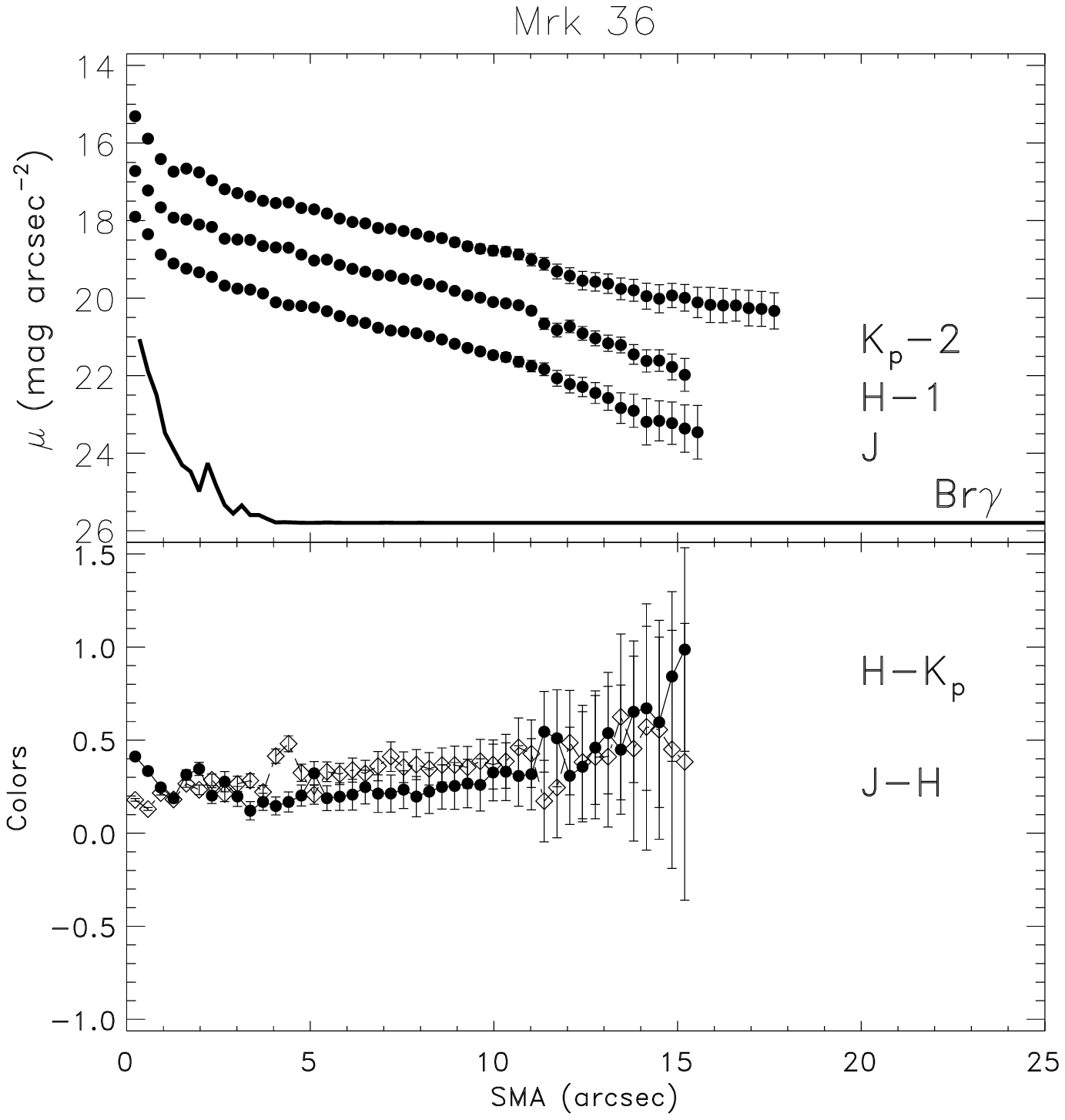}
\includegraphics[scale=0.35]{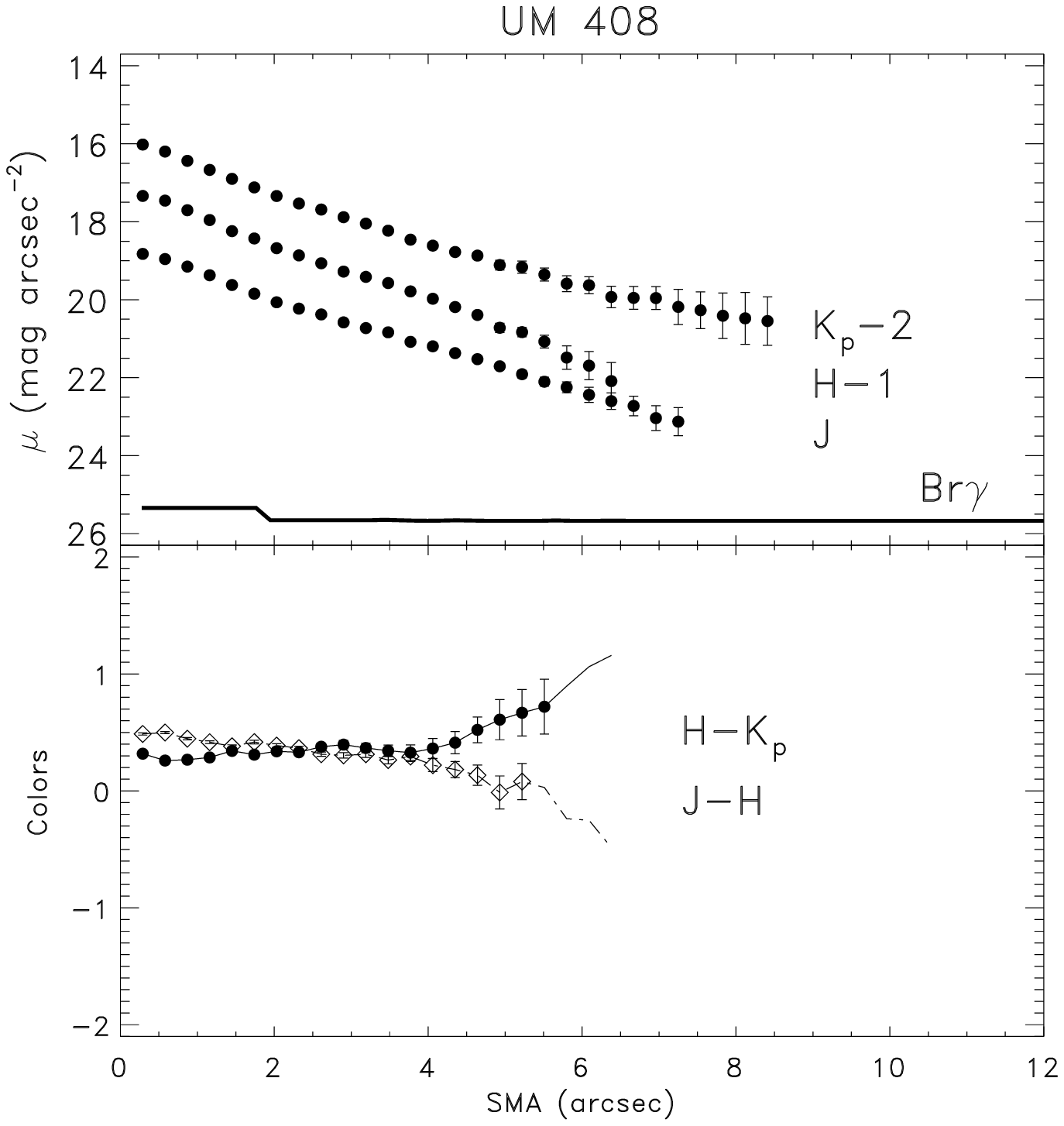}
\includegraphics[scale=0.35]{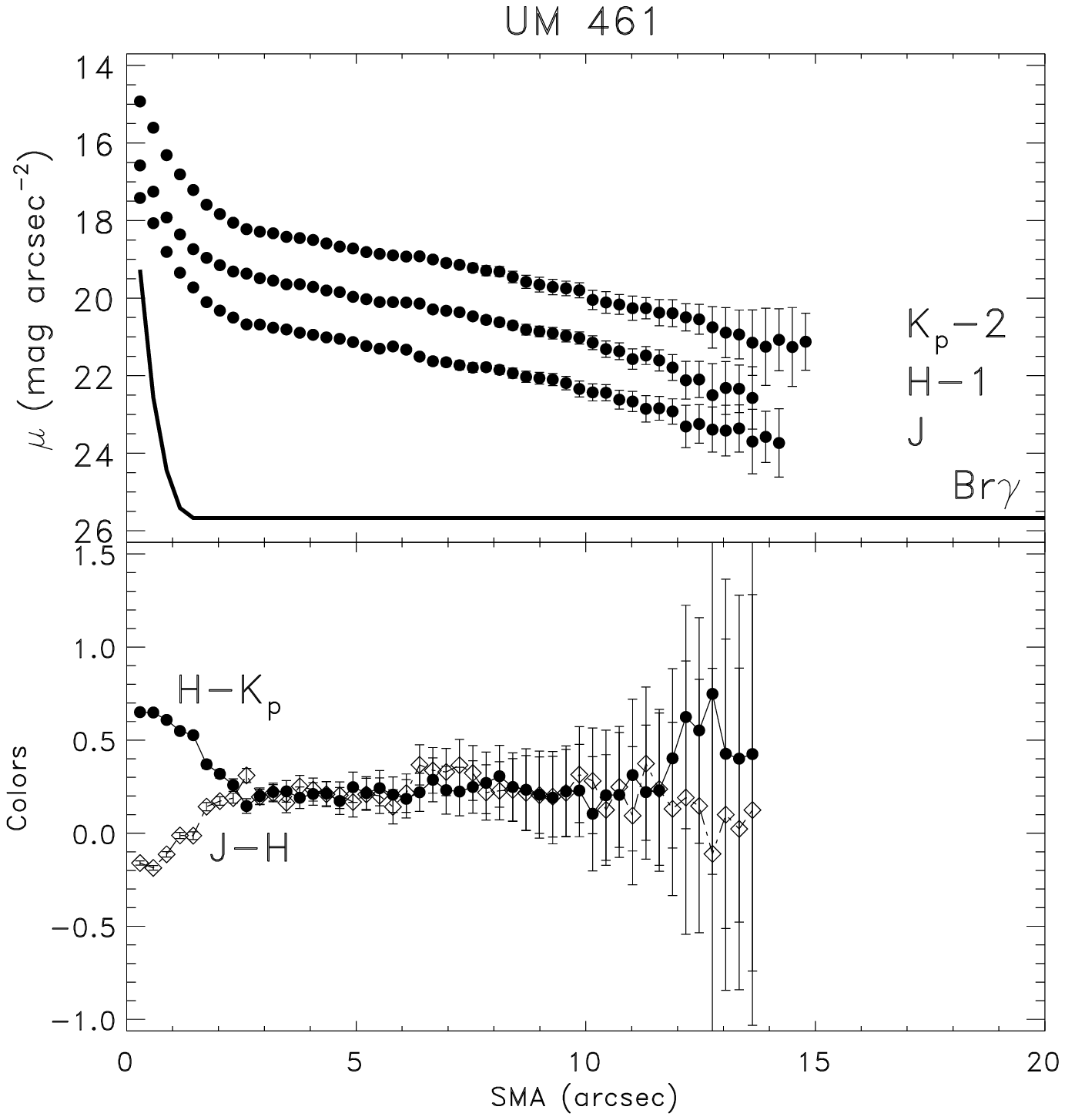}
\caption{Upper panels: surface brightness profiles of Mrk 36, UM 408, and UM 461 in J,
H, and K$_{p}$ (corrected for galactic extinction) and Br$\gamma$ in arbitrary units.
We considered pixels with 1$\sigma$ above the background.
For a better visualization, the H and K$_{p}$ profiles are shifted by -1 and -2
mag, respectively. Bottom panels:  J-H and H-K$_{p}$ color profiles.
\label{LSB}}
\end{figure}

\begin{figure*}[!ht]
\centering
\includegraphics[scale=.25]{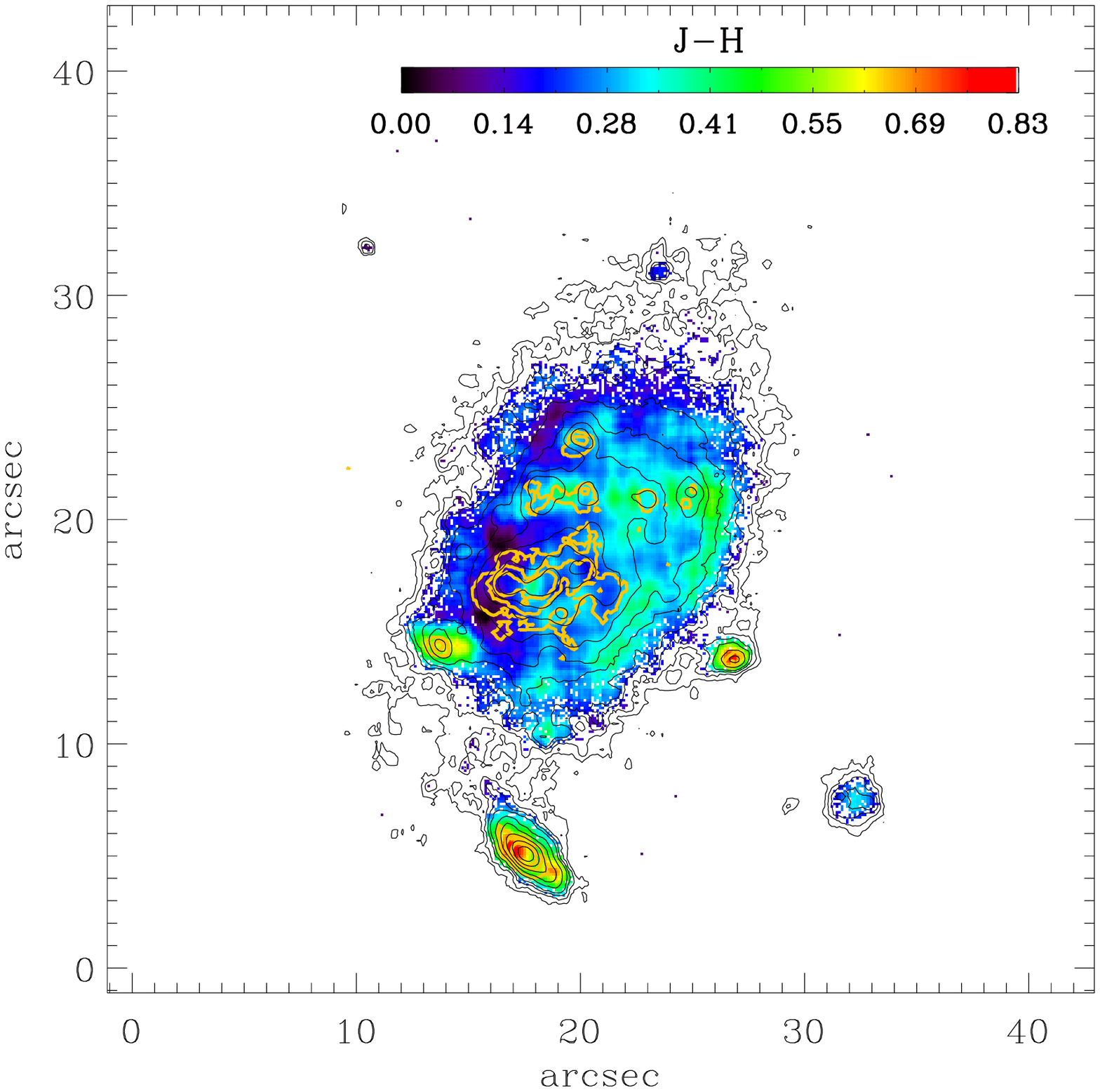}
\includegraphics[scale=.25]{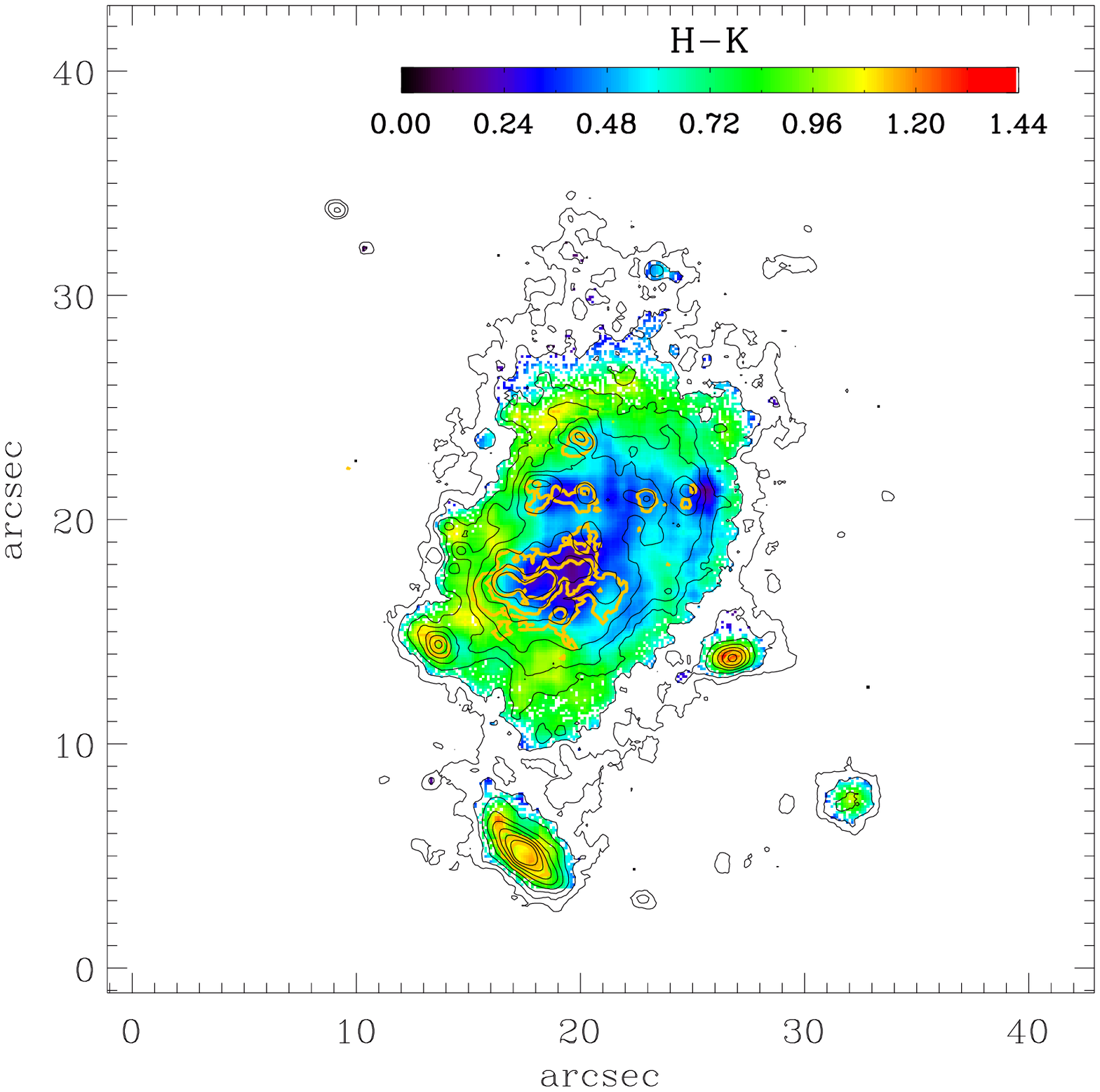}\\
\includegraphics[scale=.25]{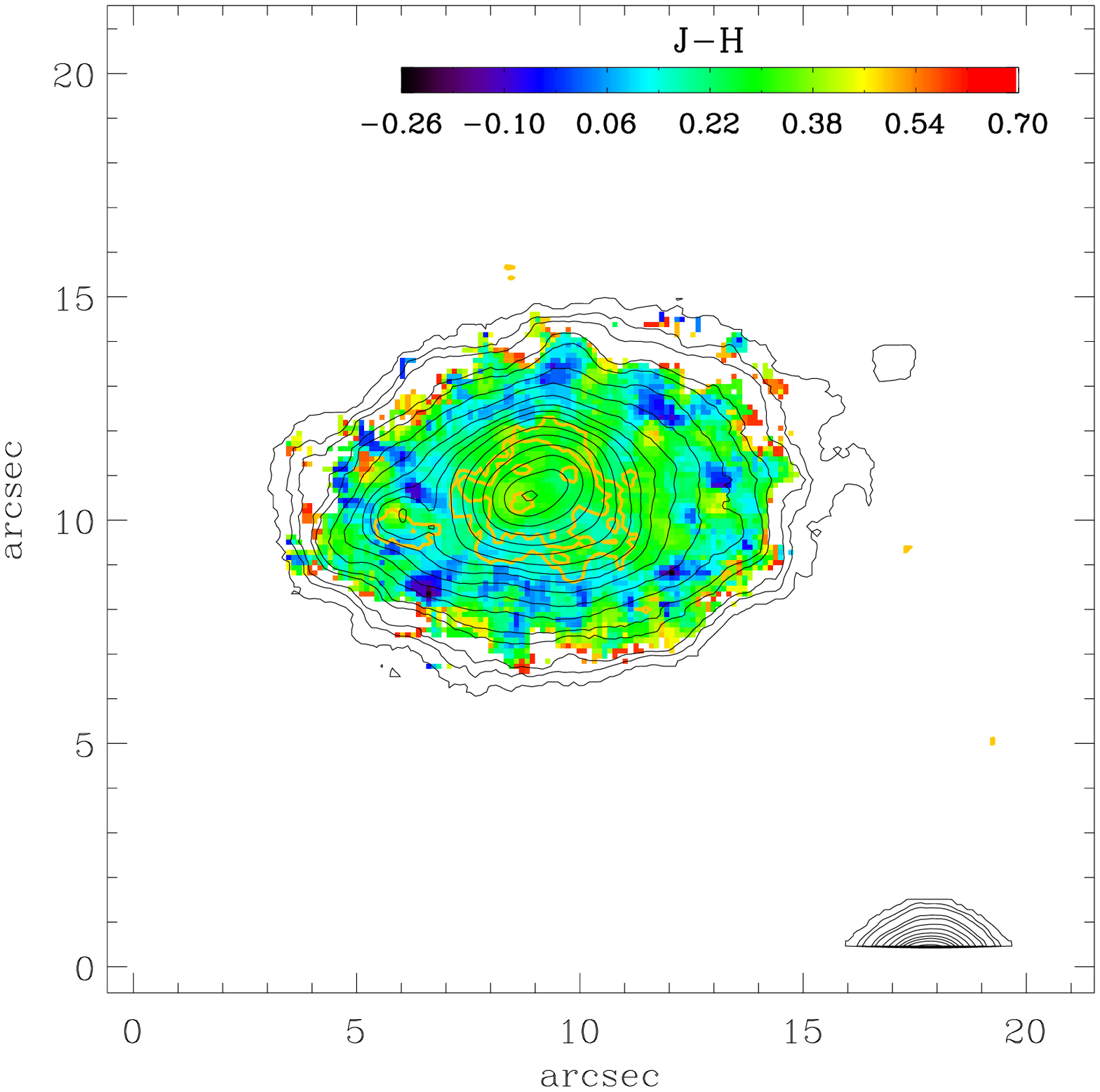}
\includegraphics[scale=.25]{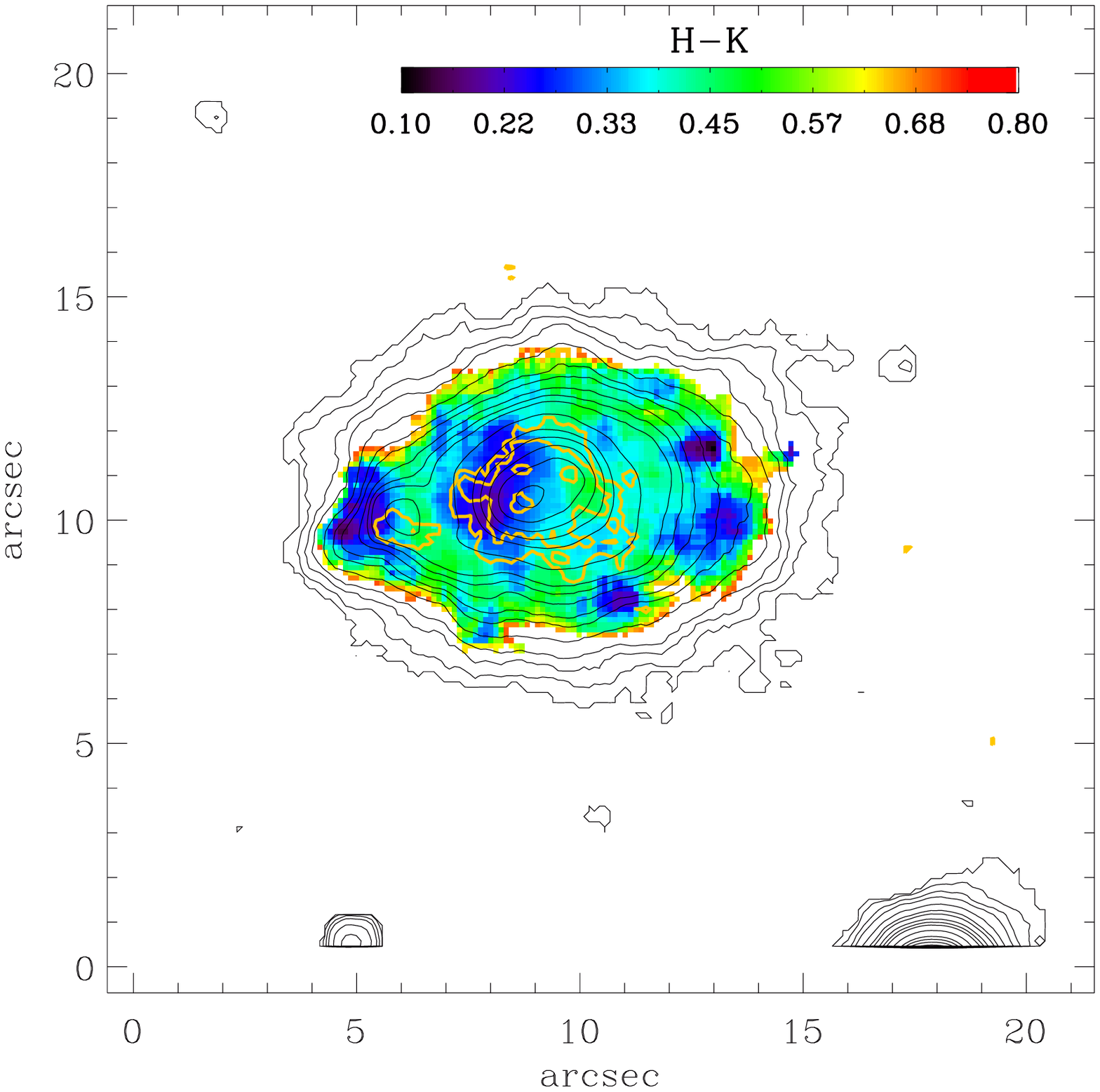}\\
\includegraphics[scale=.25]{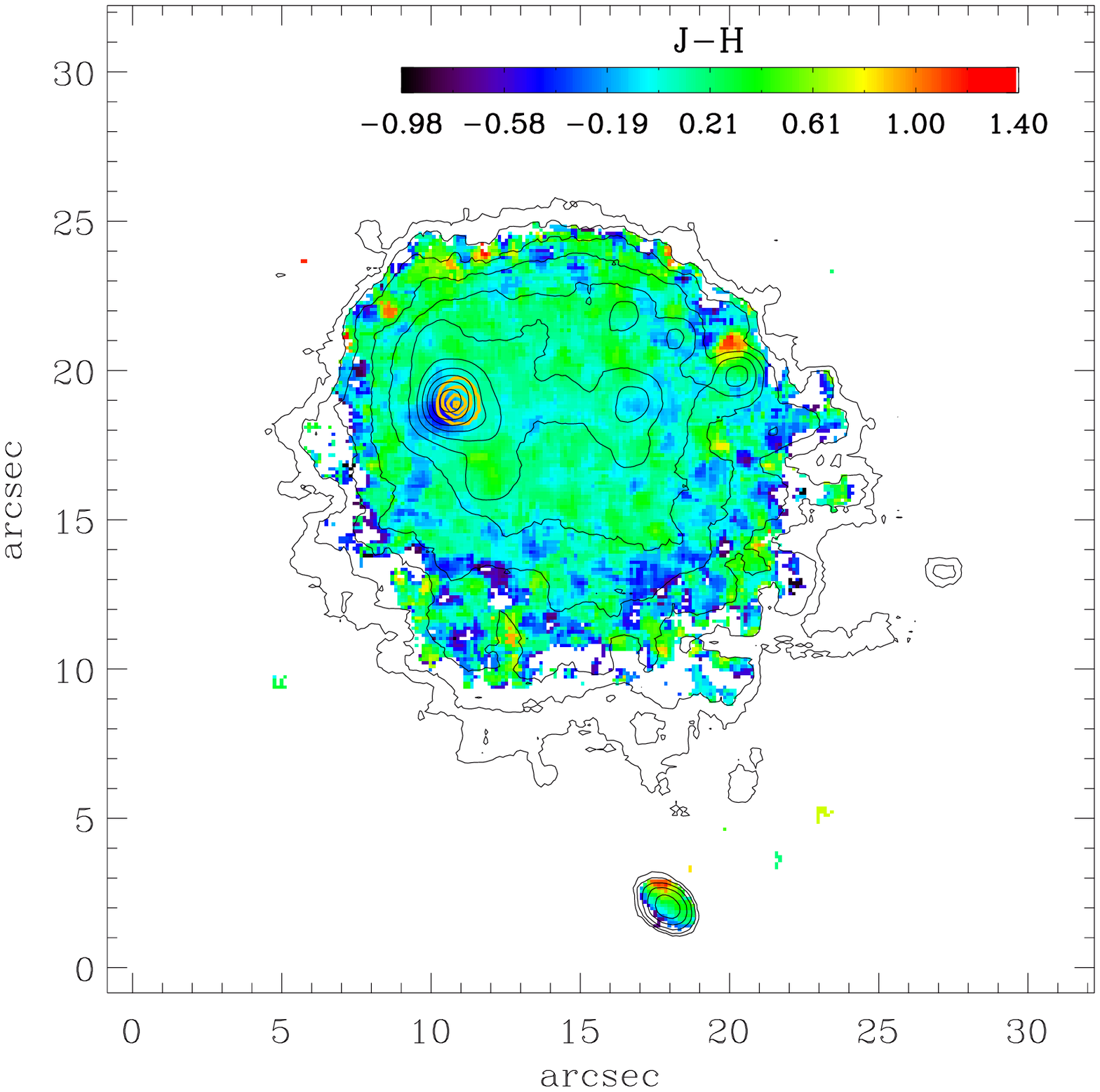}
\includegraphics[scale=.25]{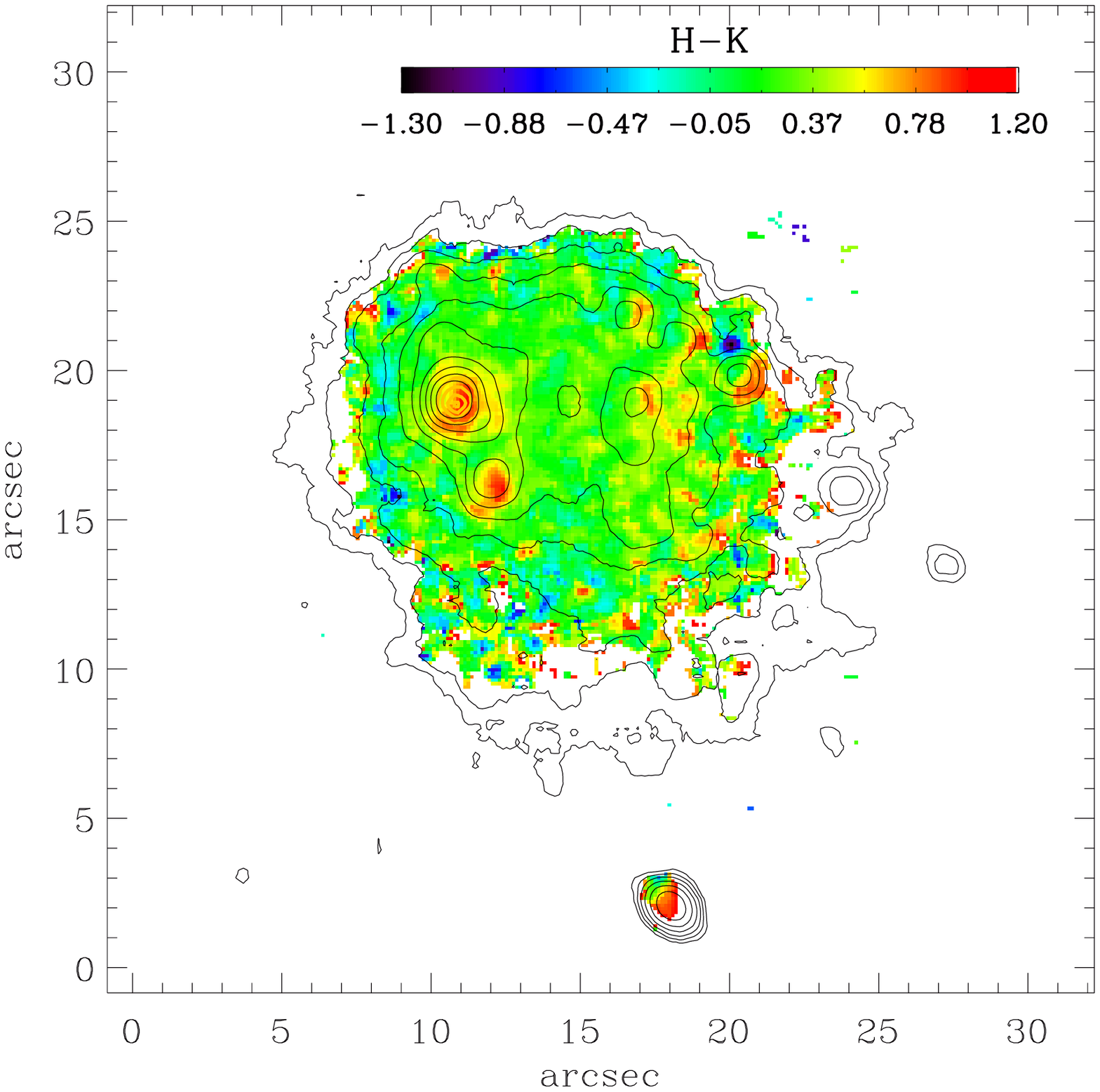}
\caption{J-H and H-K color maps of the galaxies Mrk 36 (top), UM 408 (middle), and UM 461 (bottom). 
Black contours in the J-H maps corresponds with J-band contour maps and in the H-K maps corresponds 
with K$_{p}$-band contour maps. Orange contour are from Br$\gamma$ images. 
\label{COLORS_galaxies}}
\end{figure*}

Figure~\ref{LSB} shows the surface brightness profiles (upper panels) of the isophotal mean intensity
of each galaxy in the sample,
considering pixels with 1$\sigma$ above the background, for the J, H,
K$_{p}$ and Br$\gamma$ filters. Br$\gamma$ is in arbitrary units.
We measure the fainter level limits to be $\mu_{J}\simeq$23.50, 23.10 and 23.50 mag arcsec$^{-2}$, 
$\mu_{H}\simeq$23.00, 23.10 and 23.40 mag arcsec$^{-2}$, 
and $\mu_{K_{p}}\simeq$22.30, 22.50 and 23.00 mag arcsec$^{-2}$ for Mrk 36, UM 408, and UM 461, respectively.
The J-H and H-K$_{p}$ color profiles  are shown in the lower panels of Figure~\ref{LSB}. 
We note that the color profiles show relatively constant underlying values at intermediate radii with
a color gradient at large radii, indicating that 
their stellar populations must be fairly homogeneous in the main body of the galaxy with the presence 
of a old stellar population at large radii. 
In Figure \ref{COLORS_galaxies} we show the pixel-to-pixel color map
of the galaxies in order to compare the color spatial distribution 
of these galaxies with the profiles previously obtained. 
Again, we note that the color spatial distributions are relatively constant through
the main body of the galaxies.
The surface photometry of the galaxies were corrected only for Galactic extinction
using the relation with R$_{V}$=A$_{V}$/E(B-V)=3.1 \citep{C89} and adopting the values 
from the extinction map of \citet{S98} showed in Table \ref{observation_log} Column (7). 
We did not apply any smoothing procedure to our data in order to
obtain the surface brightness at large radii. 

We see that the outer part of the light profiles in Figure~\ref{LSB} are well 
represented by an exponential model \citep[e.g.,][]{P96,T97a,C03,N03,N05}
thus other functions are not necessary. 
We can express this profile in terms of the surface
brightness, so we can find that $\mu(r)$=$\mu_{o,\lambda}$+(1.086/$\alpha_{\lambda}$)r, 
where $\mu_{o,\lambda}$ corresponds to the central surface brightness and $\alpha_{0,(J,H,K_{p})}$ the scale length
with $\lambda$=J, H and K$_{p}$.
Column (1) of Table \ref{LSB_EXPO} gives the name of
the galaxies. Columns (2), (3) and (4) give the 
$\mu_{0,(J,H,K_{p})}$, $\alpha_{0,(J,H,K_{p})}$ parameters and
m$_{LSB,(J,H,K_{p})}$ magnitudes of the LSB component described below in \S~ \ref{LSB_component}, respectively. 
These parameters will be used in \S~\ref{LSB_component} in order to calculate the integrated 
properties of the LSB component.

\subsubsection{The integrated properties of the LSB component}\label{LSB_component}

The total apparent magnitude of the LSB component can be obtained using that 
m$_{LSB}$ = $\mu_{0}$ - 5 log($\alpha$) - 1.995 - 2.5 log(1-$\varepsilon$), where 
$\varepsilon$ correspond to the ellipticity derived from the surface brightness profile 
(with constant values of $\varepsilon$=0.40, 0.30 and 0.33 for the galaxies Mrk 36, UM 408 and UM 461, respectively).
So,  we calculated the J-H and H-K colors of the LSB component from the results of the exponential fits. 
The colors of the LSB component in Mrk 36 are similar to those obtained previously in the
literature by \citet{C03} with J-H=0.37 and
H-K$_{s}$=0.28 from exponential fits. For UM 461, \citet{N03} found J-H and H-K$_{s}$ colors, of the LSB component, that 
disagree and agree with our values at 1$\sigma$ level, respectively. 
The reason for this disagreement could be the uncertainties in the sky estimation and/or 
the aperture differences in both studies causing
the surface brightness profile obtained by \citet{N03} to be deeper with the radius compared with our profiles, 
thus producing higher slopes of the profiles at large radii.  
In summary, the LSB component of our sample of galaxies show blue J-H colors and red H-K that suggest  
a photometrically dominant stellar population of ages $\gtrsim$10$^{8}$ yr.  
Our interpretation agrees with the age estimates given by \citet{R00} and \citet{W04} for a sample of
H\,{\sc ii} galaxies (in which they include UM 408 and UM 461 in their analysis) by
means of spectral population synthesis.

Finally, we estimated the stellar mass (M$_{\ast}$) of the LSB component for each galaxy
using the M$/$L relationship for the H-band \cite[see][]{LS10}. We found that the stellar mass for our sample of galaxies
are log(M$_{\ast}$)= 7.89, 8.33, and 7.88 M$_{\odot}$ for Mrk 36, UM 408, and UM
461, respectively. 
These stellar masses are typical for BCD galaxies calculated
using different photometric bands, and dominated by the contribution of
the intermediate to old stellar population \citep[e.g.,
3$\times$10$^{7}$M$_{\odot}$ for Mrk 36;][]{A09}. 

\section{Discussion}\label{discussion}

Star formation occurs when the local molecular gas density exceeds a certain
threshold \citep[see][and references therein]{L08}, given first possibly by collisions of gas clouds, 
due to turbulence or gravity, and then triggered by the action of
massive star evolution, e.g. stellar winds from star clusters and SNe.
In many cases, the current star-formation may be triggered by external agents, as in the
case of the tidal forces by a neighboring galaxy or mergers. However, for a fraction of isolated, 
less luminous and compact H\,{\sc ii} galaxies, this star-formation activity may occur solely by
internal processes. 

The SF activity may form a fraction 
of bound star clusters that will survive their infant mortality and evolve.   
\citet{B08} has defined the value of the present cluster formation efficiency $\Gamma$ in a host galaxy  as CFR/SFR, 
where CFR = M$_{tot}$/$\Delta$t is the present cluster formation rate at which the galaxy produces a total cluster mass 
M$_{tot}$ in a given age interval $\Delta$t. On the other hand, \citet{G10} found a correlation between the value of
the present cluster formation efficiency and the SF density of the host galaxy in a sample of starburst galaxies, 
$\Gamma$(\%)=29$\Sigma^{0.24}_{SFR}$ M$_{\odot}$ yr$^{-1}$ kpc$^{-2}$, where 
$\Sigma_{SFR}$ is the total SFR per unit of area. 

The first step in the determination of the star cluster formation efficiency
in our sample of galaxies is the determination of the total SF rate (SFR). 
We calculate the total SFR\footnote{We multiplied the SFR by a factor 0.67 obtained from
the comparison between the SFR of our galaxies using the \citet{K98} and \citet{C07} relationships 
This due to differences in the stellar IMF assumptions given that the \citet{K98} relationship is based on a Salpeter IMF and 
the models which we adopted are based on a Kroupa IMF} 
using the \citet{K98} relationship for L(H$\alpha$). 
We use the integrated H$\beta$ flux obtained
using continuum-free emission line broad band images, with
L(H$\alpha$)/L(H$\beta$)=2.87, by \cite{L07} for Mrk 36 and UM 461, and integrated integral field unit H$\alpha$ flux obtained 
by \cite{L09} for UM 408. 
So we find that the total SFR(H$\alpha$)$\simeq$ 0.078 M$_{\odot}$ yr$^{-1}$ in Mrk 36, 0.017 M$_{\odot}$ yr$^{-1}$ in UM 408, 
and 0.085 M$_{\odot}$ yr$^{-1}$ in UM 461.
The SFR in our sample of galaxies is very low compared with other dwarf galaxies, such as Haro 11
\citep[22  M$_{\odot}$ yr$^{-1}$;][]{A10} and NGC 1519 \citep[0.3626 M$_{\odot}$ yr$^{-1}$;][]{G10} and 
in fact with respect to more irregular and luminous H\,{\sc ii}/BCD galaxies, such as Tol 9 with a
SFR(H$\alpha$)= 1.82 M$_{\odot}$ yr$^{-1}$ \citep{LS10}. 
Assuming a starburst area of $\sim$2 kpc$^2$ in Mrk 36, $\sim$8.16 kpc$^2$ in UM 408 
and $\sim$1.62 kpc$^2$ in UM 461, we found that 
the integrated SFR per unit of area in Mrk 36 is $\Sigma_{SFR}$=0.039 M$_{\odot}$ yr$^{-1}$ kpc$^{-2}$, in UM 408 is 
$\Sigma_{SFR}$=0.002 M$_{\odot}$ yr$^{-1}$ kpc$^{-2}$ 
and in UM 461 is $\Sigma_{SFR}$=0.052 M$_{\odot}$ yr$^{-1}$ kpc$^{-2}$, respectively.
We can now calculate the expected value of $\Gamma$ using 
the correlation found by \citet{G10}. 
Thus, we obtain that the cluster formation efficiency is $\sim$13\% in Mrk 36,
$\sim$7\% in UM 408 and $\sim$14\% in UM 461, respectively. 
Hence, the star cluster efficiency in our low luminosity H\,{\sc ii} galaxies 
is approximately $\sim$10\%. This current SF efficiency is lower than the ones found 
in more luminous BCD galaxies as Haro 11 \citep{A10} with an efficiency 
of $\sim$38\%. 

Using the properties of our detected star clusters/complexes showed in Table \ref{clusters_data}, 
we calculate a total cluster/complex mass, by summing the masses,
of 54.50$\times$10$^{4}$M$_{\odot}$ in Mrk 36 
(with a total mass in clusters $\gtrsim$10$^4$M$_{\odot}$ 
of 44.69$\times$10$^{4}$ M$_{\odot}$), 13.81$\times$10$^{5}$M$_{\odot}$ in UM 408 and 
43.09$\times$10$^{4}$M$_{\odot}$ in UM 461. 
So, we calculate that the fraction of the total cluster mass 
with respect to the LSB host galaxy mass in Mrk 36, considering
our complete range in ages, is equal to $\sim$0.007, $\sim$0.006 in UM 408 and $\sim$0.006 in UM 461. 
This implies that the inferred total current SF mass is of the order of $\lesssim$1\% of the underlying galaxy mass,
in agreement with the estimate of \cite{W04} that the past history of SF in the galaxies
were more active than the present one.
Additionally, we calculate the SFR per unit of area for the individual star clusters/complexes using 
our measured Br$\gamma$ emission. From these, we obtain that 
in the brightest regions (clusters \#1 and \#5 in Mrk 36 and complex \#5 in UM 461) 
the SFR per unit of area is compatible with the values observed 
in starburst galaxies with values $>$0.1 M$_{\odot}$ yr$^{-1}$ kpc$^{-2}$ \citep[][and references therein]{Ba05}.  
While, the majority of the star clusters/complexes have values of the order of $\sim$0.01 M$_{\odot}$ yr$^{-1}$ kpc$^{-2}$. 
If we use only the total mass in clusters $\gtrsim$10$^4$M$_{\odot}$ to calculate SF efficiency, we obtain that
in  Mrk 36 for an age of 20 Myr our result agree, within the uncertainties, with the one
obtained using the relationship of \cite{G10}. However, in the case of 
UM 408 and UM 461 the value of $\Gamma$ is extremely high, indicating that in these cases the star forming complexes
are not resolved into individual clusters, resulting in a overestimation of the $\Gamma$ parameter.

We can also calculate the gas consumption time scale ($\tau_{gas}$) 
or how long it would take before all the gas 
in the galaxies will be consumed at the current SFR. The consumption time scale is defined as
the ratio between the available gas and the current SFR, $\tau_{gas}$ = M$_{gas}$/SFR. Assuming 
the total amount of gas M$_{gas}$ = M$_{HI}$+M$_{He}$+M$_{H_2} \approx$ 2$\times$M$_{HI}$ \citep{L05},
we obtain that $\tau_{gas} \sim$ 0.5 Gyr in Mrk 36 \citep[with M$_{HI}$=2.0$\times$10$^{7}$M$_{\odot}$;][]{TM81,BA04}, 
$\tau_{gas} \sim$ 77 Gyr in UM 408 \citep[with M$_{HI}$=6.53$\times$10$^{8}$M$_{\odot}$;][]{S02}
and $\tau_{gas} \sim$ 2 Gyr in UM 461 \citep[with M$_{HI}$=0.98$\times$10$^{8}$M$_{\odot}$;][]{S00}.
The consumption timescales in Mrk 36 and UM 461 are significantly less than a Hubble time and 
comparable with the times observed in spiral galaxies \citep{K94}. So the SF 
cannot be sustained for the entire history of the galaxies,
which indicates that these objects undergo a few or several short bursts of SF. 
While UM 408 has a gas consumption timescale longer than a Hubble time, indicating that likely the SF
is relatively constant through the history of the galaxy or the HI halo may not be spatially available for the
current SF. 

Our findings also seem to indicate that the SF mode in our sample 
of low luminosity H\,{\sc ii} galaxies is clumpy, similar to other dwarf galaxies (e.g., NGC 1569 and SBS 0335-052). 
These complexes or star-forming knots are formed by a few massive star clusters 
with masses $\gtrsim$10$^{4}$M$_{\odot}$ and high SFR per unit of area.
\citet{M09} found that the observed trends in the number and mass of the SF regions, 
in a sample of dwarf galaxies, is independent of the local environment and even the surrounding galaxy mass
and is given by a mass function which stochastically favored SF in clusters \citep{A10}.
\citet{B02} show that SSCs may require special (or fortunate) circumstances to
form in dwarf galaxies, but when they do, they are very massive
($\gtrsim$10$^{4-5}$M$_{\odot}$) and form clumps or groups of similar ages.
This suggests that these clumps are likely formed in localized regions of high
pressure triggered by large scale ambient gravitational instabilities,
given that in dwarf galaxies most of the ISM is at low pressure \citep{E00}.
The lack of external perturbers in the most compact and isolated galaxies indicates 
that an additional mechanism other than tidal interactions must be considered 
to explain this current SF activity. 
This mechanism may be related to the overall physical conditions of
the ISM, particularly the gas surface densities, in conjunction with stochastic
effects, that allow SF to take place.
Alternatively, the low current SFR implies that a burst or a triggering is not necessary, simply 
that the SFR has been relatively constant.
However, tidal interactions or mergers are likely the primary agent to trigger 
the current SF in luminous and more disturbed H\,{\sc ii} galaxies
as suggested by their morphology  \citep[e.g.,][]{T95,L07,LS08}.

\section{Conclusions}\label{conclusions}

In this paper, a sample of three H\,{\sc ii} galaxies (Mrk 36, UM 408, and UM
461) has been analyzed in order to study their 
stellar populations (star cluster complexes and the underlying host galaxy or LSB component) 
using new near-IR high spatial resolution images obtained on the
Gemini North telescope. In our analysis we used models that include 
the contribution of stellar continuum, nebular continuum and emission lines.
Our conclusions can be summarized as follows:

\begin{enumerate}

\item 
The presence of nebular continuum and emission lines in the near-IR 
produces an excess in the observed SED in young star cluster/complexes. 
This excess, can have a large impact in the
inferred properties of the star clusters in H\,{\sc ii}/BCD galaxies 
and models that include this effect are the most appropriate in the study of 
young stellar population with ages $\lesssim$6 Myr. 

\item We found that the star cluster population in Mrk 36 shows masses of 
$\sim$10$^{4-5}$M$_{\odot}$ with a few detected star clusters with masses 
of $\sim$10$^3$M$_{\odot}$ distributed in the main body of the galaxy. 
The star cluster complexes in UM 408 and UM 461 have masses from $\sim$10$^{4}$M$_{\odot}$ to
$\sim$10$^{6}$M$_{\odot}$. The age distribution shows that the detected star clusters/complexes are very 
young with ages of a few Myr. Two likely old star clusters with colors consistent with ages $>$10-100 Myr 
have been detected in Mrk 36. 
The fraction of recent SF in bound clusters and/or complexes more massive than 10$^{4}$M$_{\odot}$
is compared to the current total SFR is about 10\% in our sample of galaxies.

\item 
The spatial distribution and ages of the star
cluster/complex population seems to indicate 
that SF is clumpy and simultaneous. 
We propose that the current SF activity in our sample of low luminosity H\,{\sc ii}
galaxies is triggered by some internal mechanism 
instead of tidal interactions.
This mechanism of SF may be related to the overall physical
conditions of the ISM that produce the increase 
of surface densities in conjunction with stochastic effects 
within a time scale comparable to the mean age differences of the massive star cluster complexes.

\item The LSB component of our sample of galaxies have near-IR colors
representative of evolved stellar population of at least $\gtrsim$10$^{8}$ yr. 
We found that the stellar mass of this component for our sample of galaxies
are log(M$_{\ast}$)= 7.89, 8.33, and 7.88 M$_{\odot}$ for Mrk 36, UM 408, and UM
461, respectively. The fraction of the total cluster mass
with respect to the LSB hosting galaxy mass in our sample of galaxies, considering
our complete range in ages, is less than 1\%.

\end{enumerate}
 
\acknowledgments
P.L. is supported by a Post-Doctoral grant (SFRH/BPD/72308/2010), 
funded by FCT/MCTES (Portugal) and POPH/FSE (EC).
P.L. would like thank to Polychronis Papaderos and Damian
Fabbian for their comments and very useful discussions. 
A.N. acknowledges support by the projects:
AYA2007-67965-C03-02, AYA2010-21887-C04-01 and
Consolider-Ingenio 2010 CSD2006-00070 “First Science with GTC”,
of the spanish MICINN.
We thank Ralf Kotulla for providing us with the GALEV models used
in this work. We also thank the anonymous referee for numerous useful comments and 
suggestions which led to the overall improvement of this paper. 
Based on observations obtained at the Gemini Observatory, which is
operated by the Association of Universities for Research in Astronomy, Inc., under a
cooperative agreement with the NSF on behalf of the Gemini partnership: the
National Science Foundation (United States), the Science and Technology
Facilities Council (United Kingdom), the National Research Council (Canada),
CONICYT (Chile), the Australian Research Council
(Australia), Minist\'erio da Ci\^encia e Tecnologia (Brazil) and Ministerio de
Ciencia,
Tecnolog\'{\i}a e Innovaci\'on Productiva (Argentina). Program ID: GN-2005B-Q-42.
This research has made use of the NASA/IPAC Extragalactic Database (NED) which
is operated by the Jet Propulsion laboratory, California Institute of
technology, under contract with the National Aeronautics and Space
Administration.

{\it Facilities:} \facility{Gemini:North (NIRI)}

\clearpage

\begin{deluxetable}{cccccccccccc}
\tabletypesize{\scriptsize}
\rotate
\tablecolumns{11}
\tablecaption{Sample data and NIRI observations. \label{observation_log}}
\tablewidth{0pt}

\tablehead{Object & $\alpha$& $\delta$ & $Vel.$ & 12+log(O/H) & c(H$\beta$)&
E(B-V)$_{Gal}$& Date of  &  Filter & Exposure &  Air mass & C$_{\lambda}$\\
       &  (J2000)& (J2000)  & (km/s) &             &            &   &
Observation&         & Time (s) &        & (mag) \\
 (1)   &    (2)  &   (3)    &  (4)   & (5)         &    (6)     & (7)  &  (8)   
   &   (9)   &   (10)  &   (11) & (12)\\}
\startdata
Mrk 36  &11:04:44.0 &+29:07:48 & 646 & 7.81\tablenotemark{a}& 
0.02\tablenotemark{a} & 0.031& 2005-11-24 &   &   & \\
        &           &      & (12.95 Mpc)&  &            &
&&J          &6$\times$(3$\times$60)   & 1.37 & 23.93$\pm$0.05\\
        &           &          &     &  &            & &&H         
&6$\times$(4$\times$65)   & 1.27 & 24.01$\pm$0.05\\
        &           &          &     &  &            & &&K$_{p}$   
&12$\times$(3$\times$75)  & 1.15 & 23.57$\pm$0.06\\
        &           &          &     &  &            & &&Br$\gamma$
&6$\times$(3$\times$40)   & 1.08\\
UM 408  &02:11:23.4 &+02:20:30 & 3598& 7.87\tablenotemark{b} &
0.93\tablenotemark{b} & 0.037& 2005-08-02 &   &   &  \\
        &           &      &  (45.70 Mpc)   &  &         
  & &&J          &6$\times$(3$\times$60)   & 1.26 & 23.91$\pm$0.05\\
        &           &          &     &  &            & &&H         
&6$\times$(4$\times$65)   & 1.19 & 23.97$\pm$0.03\\
        &           &          &     &  &            & &&K$_{p}$   
&12$\times$(3$\times$75)  & 1.10 & 23.49$\pm$0.04\\
        &           &          &     &  &            & &&Br$\gamma$
&6$\times$(3$\times$40)   & 1.06\\
UM 461  &11:51:33.1 &-02:22:22 & 1039& 7.78\tablenotemark{a} &
0.12\tablenotemark{a} & 0.018& 2005-12-29 &   &   &  \\
        &           &   & (19.20 Mpc)    &  &            &
&&J          &6$\times$(3$\times$60)   & 1.62 & 23.95$\pm$0.07\\
        &           &          &     &  &            & &&H         
&6$\times$(4$\times$65)   & 1.46 & 24.05$\pm$0.05\\
        &           &          &     &  &            & &&K$_{p}$   
&12$\times$(3$\times$75)  & 1.28 & 23.54$\pm$0.02\\
        &           &          &     &  &            & &&Br$\gamma$
&6$\times$(3$\times$40)   & 1.18\\
\enddata
\tablenotetext{a}{\citet{I98}}
\tablenotetext{b}{\citet{L09}}
\tablecomments{Column (1) galaxy name. Columns (2) and (3) 
$\alpha$ and $\delta$ coordinates (J2000), respectively. Column (4) 
heliocentric velocity
(vel.) and the 3K CMB corrected distance from NED. Columns (5) and (6) 
oxygen abundance and extinction adopted in this work. Column (7) 
date of observation. Column (8) Galactic extinction obtained from
the extinction maps of \citet{S98}. Column (9) filter
used in each observation and finally columns (10) and (11) exposure
time in s and the mean air mass of each observation, respectively. Column (12) shows the 
instrumental zero points C$_{\lambda}$ with $\lambda$= J, H and K$_{p}$ obtained for each run of
observation.
}
\end{deluxetable}

\clearpage

\begin{table*}
\begin{center}
\caption{Integrated magnitudes of our sample of galaxies.\label{mag_int}}
\begin{tabular}{cccc}
\tableline\tableline
  Object &    J           &  H           & K$_{p}$    \\
         &    (mag)       &  (mag)       & (mag)      \\
  (1)    &     (2)        &      (3)     &      (4)         \\
\tableline
Mrk 36   & 14.46$\pm$0.05  &  14.23$\pm$0.05  &  13.70$\pm$0.06 \\

UM 408   & 15.94$\pm$0.07  &  15.72$\pm$0.06 &  15.28$\pm$0.05\\

UM 461   & 15.04$\pm$0.09  & 14.73$\pm$0.07 & 14.53$\pm$0.05\\
\tableline
\end{tabular}
\end{center}
\end{table*}

\clearpage

\begin{deluxetable}{cccccccccc}
\tablewidth{0pt}
\tabletypesize{\scriptsize}
\rotate
\tablecaption{Observed aperture photometry in the near-IR bands J, H and K$_{p}$
and measured properties of the stellar clusters. \label{clusters_data}}
\tablehead{
Name  &Cluster& Aperture radii  &   J   &   H   & K$_{p}$ & E(B-V) & Age & Mass
& Log L(Br$\gamma$)\\
      &       & (arcsec)        & (mag) & (mag) & (mag)   &  (mag) & (Myr) &
($\times$10$^{4}$M$_{\odot}$) & (erg s$^{-1}$)\\
  (1) & (2)   & (3)             & (4)   & (5)   & (6)     &  (7)   & (8) & (9) &
(10)}
\startdata
\dataset[Mrk 36]{Mrk 36} & \\
       &  1$^{\dagger}$ & 0.44 & 18.64$\pm$0.05 & 18.58$\pm$0.05 & 18.09$\pm$0.06 & 0.00 -- 0.40 &1.99 -- 2.8 & 5.46 -- 2.53 & 36.256\\
       &  2             & 0.39 & 20.01$\pm$0.05 & 20.13$\pm$0.05 & 19.44$\pm$0.06 & 0.00 -- 0.35 &1.25 -- 1.1 & 1.73 -- 0.64 & $\cdots$\\
       &  3$^{\dagger}$ & 0.28 & 19.98$\pm$0.05 & 19.58$\pm$0.05 & 19.40$\pm$0.06 & 0.50 -- 0.75 &10.11-- 8.6&  4.08 -- 3.62  &35.268\\
       &  4$^{\dagger}$ & 0.30 & 19.88$\pm$0.05 & 19.51$\pm$0.05 & 19.24$\pm$0.06 & 1.00 -- 1.00 &5.07 -- 5.5&  3.49 -- 1.77  &35.818\\
       &  5$^{\dagger}$ & 0.29 & 19.99$\pm$0.05 & 19.68$\pm$0.05 & 19.46$\pm$0.06 & 0.95 -- 0.75 &6.09 -- 5.5&  3.56 -- 1.33  &35.912\\
       &  6$^{\dagger}$ & 0.37 & 19.84$\pm$0.05 & 19.45$\pm$0.05 & 19.19$\pm$0.06 & 1.00 -- 1.00 &5.07 -- 5.5&  3.66 -- 1.86  &35.314\\
       &  7$^{\dagger}$ & 0.29 & 19.87$\pm$0.05 & 19.52$\pm$0.05 & 19.23$\pm$0.06 & 1.00 -- 0.90 &4.89 -- 5.0&  3.81 -- 1.64 &35.538\\
       &  8$^{\dagger}$ & 0.40 & 18.83$\pm$0.05 & 18.62$\pm$0.05 & 18.45$\pm$0.06 & 0.60 -- 0.50 &5.62 -- 5.9&  6.78 -- 3.60 &35.072\\
       &  9             & 0.24 & 20.74$\pm$0.05 & 20.54$\pm$0.05 & 20.21$\pm$0.06 & 0.20 -- 0.65 &2.95 -- 4.2&  0.97 -- 0.62 & $\cdots$\\
       & 10             & 0.28 & 19.99$\pm$0.05 & 19.61$\pm$0.05 & 19.57$\pm$0.06 & 0.00 -- 0.00 &$\sim$1000 -- 19.3 &45.62 -- 4.50  & 35.523 \\
       & 11             & 0.29 & 19.74$\pm$0.05 & 19.62$\pm$0.05 & 19.50$\pm$0.06 & 0.30 -- 0.25 &5.43 -- 6.5&  2.18 -- 1.70 & $\cdots$\\
       & 12$^{\dagger}$ & 0.24 & 20.48$\pm$0.05 & 20.28$\pm$0.05 & 20.19$\pm$0.06 & 0.20 -- 0.25 &7.85 -- 8.3&  1.18 -- 1.44 &34.273\\
       & 13             & 0.29 & 20.57$\pm$0.05 & 20.43$\pm$0.05 & 20.06$\pm$0.06 & 0.00 -- 0.45 &2.78 -- 4.0&  0.99 -- 0.70 & $\cdots$\\
       & 14             & 0.32 & 20.62$\pm$0.05 & 20.37$\pm$0.05 & 20.05$\pm$0.06 & 0.45 -- 0.75 &3.67 -- 4.3&  1.37 -- 0.71 & $\cdots$ \\
       & 15             & 0.43 & 20.97$\pm$0.05 & 20.83$\pm$0.05 & 20.21$\pm$0.06 & 0.30 -- 0.75 &1.25 -- 1.0&  0.91 -- 0.36 & $\cdots$\\
       & 16             & 0.38 & 21.10$\pm$0.05 & 20.97$\pm$0.05 & 20.50$\pm$0.06 & 0.00 -- 0.65 &1.64 -- 3.5&  0.60 -- 0.42 & $\cdots$\\
       & 17$^{\dagger}$ & 0.56 & 19.39$\pm$0.05 & 19.10$\pm$0.05 & 18.82$\pm$0.06 & 0.75 -- 0.80 &4.46 -- 4.8&  5.52 -- 2.26 &35.374\\
       & 18             & 0.52 & 19.59$\pm$0.05 & 19.54$\pm$0.05 & 18.87$\pm$0.06 & 0.20 -- 0.65 &1.25 -- 1.0&  3.08 -- 1.21 & $\cdots$\\
       & 19$^{\dagger}$ & 0.44 & 19.80$\pm$0.05 & 19.39$\pm$0.05 & 19.21$\pm$0.06 & 0.50 -- 0.80 &10.23 -- 8.5&  4.77 -- 4.40 &33.516\\
       & 20$^{\dagger}$ & 0.40 & 19.99$\pm$0.05 & 19.70$\pm$0.05 & 19.39$\pm$0.06 & 0.70 -- 0.70 &4.16 -- 4.5&  3.03 -- 1.33 &35.497\\
       & 21             & 0.40 & 20.40$\pm$0.05 & 19.96$\pm$0.05 & 19.80$\pm$0.06 & 0.25 -- 0.55 &26.92-- 9.6& 3.39 -- 2.26 & $\cdots$\\
       & 22             & 0.27 & 21.19$\pm$0.05 & 21.05$\pm$0.05 & 20.76$\pm$0.06 & 0.00 -- 0.50 &2.95 -- 4.3& 0.54 -- 0.34 & $\cdots$\\
       & 23             & 0.41 & 20.28$\pm$0.05 & 19.92$\pm$0.05 & 19.61$\pm$0.06 & 1.00 -- 1.00 &4.57 -- 4.8& 2.90 -- 1.18 & $\cdots$\\
       & 24$^{\dagger}$ & 0.47 & 19.75$\pm$0.05 & 19.35$\pm$0.05 & 19.14$\pm$0.06 & 0.85 -- 0.95 &15.85  -- 7.9& 9.72 -- 4.79 &35.060\\
       & 25$^{\dagger}$ & 0.36 & 20.68$\pm$0.05 & 20.40$\pm$0.05 & 20.13$\pm$0.06 & 0.75 -- 0.75 &4.57  -- 4.8& 1.64 -- 0.66 &34.859\\
       & 26$^{\dagger}$ & 0.33 & 20.75$\pm$0.05 & 20.50$\pm$0.05 & 20.10$\pm$0.06 & 0.05 -- 0.85 &1.25  -- 4.1& 0.89 -- 0.77 &34.829\\
       & 27             & 0.33 & 20.89$\pm$0.05 & 20.54$\pm$0.05 & 20.61$\pm$0.06 & 0.00 -- 0.00 &$\sim$1000 -- 17.9 &17.20-- 1.61 & $\cdots$\\
       & 28$^{\dagger}$ & 0.36 & 20.39$\pm$0.05 & 20.07$\pm$0.05 & 19.74$\pm$0.06 & 0.80 -- 0.95 &4.21  -- 4.5& 2.31 -- 0.98 &35.261\\
       & 29$^{\dagger}$ & 0.40 & 19.99$\pm$0.05 & 19.76$\pm$0.05 & 19.32$\pm$0.06 & 0.10 -- 0.80 &1.25  -- 4.0& 1.87 -- 1.60 &33.584\\
       & 30             & 0.41 & 20.52$\pm$0.05 & 20.30$\pm$0.05 & 19.93$\pm$0.06 & 0.00 -- 0.70 &2.39 -- 4.1& 0.99  -- 0.85 & $\cdots$\\
       & 31$^{\dagger}$ & 0.25 & 20.11$\pm$0.05 & 19.92$\pm$0.05 & 19.66$\pm$0.06 & 0.25 -- 0.55 &3.84 -- 4.6& 1.74  -- 0.92 &34.916\\
       & 32             & 0.50 & 20.57$\pm$0.05 & 20.31$\pm$0.05 & 19.93$\pm$0.06 & 0.35 -- 0.85 &2.88 -- 4.2& 1.32  -- 0.86 & $\cdots$\\
       & 33$^{\dagger}$ & 0.26 & 20.14$\pm$0.05 & 20.06$\pm$0.05 & 19.71$\pm$0.06 & 0.00 -- 0.35 &2.88 --  4.0& 1.44 -- 0.95 &34.997\\
\hline
UM 408 & \\
       & 1             & 0.40 & 20.52$\pm$0.05& 20.05$\pm$0.03& 19.74$\pm$0.04 & 1.00 -- 1.00 & 9.66 -- 5.1 &  43.69 -- 13.48 & $\cdots$ \\
       & 2$^{\dagger}$ & 0.53 & 18.95$\pm$0.05& 18.53$\pm$0.03& 18.24$\pm$0.04 & 1.00 -- 1.00 & 5.07 -- 5.2 & 110.22 -- 54.26 & 36.087\\
       & 3$^{\dagger}$ & 0.27 & 20.49$\pm$0.05& 20.11$\pm$0.03& 19.82$\pm$0.04 & 1.00 -- 1.00 & 4.78 -- 5.1 &  28.63 -- 12.56 &35.457\\
       & 4$^{\dagger}$ & 0.33 & 20.19$\pm$0.05& 19.88$\pm$0.03& 19.44$\pm$0.04 & 0.25 -- 1.00 & 1.25 -- 4.1 &  21.88 -- 18.57 &35.495\\
       & 5$^{\dagger}$ & 0.40 & 19.77$\pm$0.05& 19.40$\pm$0.03& 18.99$\pm$0.04 & 0.90 -- 1.00 & 3.67 -- 4.2 &  53.64 -- 27.05 &35.762\\
       & 6             & 0.17 & 21.63$\pm$0.05& 21.18$\pm$0.03& 20.92$\pm$0.04 & 1.00 -- 1.00 & 7.94 -- 8.2 &   9.81 -- 12.21 & $\cdots$\\
\hline
UM 461 & \\
       &  1             & 0.66 & 19.87$\pm$0.07 & 19.63$\pm$0.05 & 19.37$\pm$0.02 & 0.60 -- 0.70 & 4.4 -- 4.7 &  6.94-- 2.85 & $\cdots$\\
       &  2$^{\dagger}$ & 0.86 & 17.34$\pm$0.07 & 17.50$\pm$0.05 & 16.86$\pm$0.02 & 0.00 -- 0.15 & 2.0 -- 1.0 & 38.74 -- 14.33 & 37.206\\
       &  3             & 0.42 & 20.00$\pm$0.07 & 19.85$\pm$0.05 & 19.39$\pm$0.02 & 0.00 -- 0.65 & 1.3 -- 3.7 &  3.74 -- 2.88& $\cdots$ \\
       &  4             & 0.62 & 19.88$\pm$0.07 & 19.44$\pm$0.05 & 18.85$\pm$0.02 & 0.85 -- 1.00 & 1.3 -- 2.9 &  8.40 -- 3.50& $\cdots$\\
       &  5             & 0.55 & 19.86$\pm$0.07 & 19.83$\pm$0.05 & 19.38$\pm$0.02 & 0.00 -- 0.30 & 2.7 -- 2.9 &  3.96 -- 1.65& $\cdots$\\
       &  6             & 0.49 & 20.57$\pm$0.07 & 20.25$\pm$0.05 & 20.00$\pm$0.02 & 1.00 -- 0.85 & 5.5 -- 5.4 &  3.98 -- 1.82& $\cdots$\\
       &  7             & 0.50 & 20.27$\pm$0.07 & 20.08$\pm$0.05 & 19.82$\pm$0.02 & 0.30 -- 0.60 & 3.9 -- 4.6 &  3.53 -- 1.79& $\cdots$\\
       &  8             & 0.57 & 20.33$\pm$0.07 & 20.05$\pm$0.05 & 19.72$\pm$0.02 & 0.65 -- 0.85 & 3.1 -- 4.4 &  3.33 -- 2.16& $\cdots$\\
       &  9             & 0.62 & 20.15$\pm$0.07 & 20.01$\pm$0.05 & 19.79$\pm$0.02 & 0.15 -- 0.40 & 4.1 -- 4.7 &  3.71 -- 1.73& $\cdots$\\
       & 10             & 0.66 & 20.14$\pm$0.07 & 19.81$\pm$0.05 & 19.52$\pm$0.02 & 0.90 -- 0.90 & 4.6 -- 4.9 &  6.65 -- 2.75& $\cdots$\\
       & 11             & 0.61 & 20.34$\pm$0.07 & 20.03$\pm$0.05 & 19.83$\pm$0.02 & 0.85 -- 0.80 & 7.2 -- 7.0 &  5.32 -- 4.10& $\cdots$\\
       & 12             & 0.53 & 20.48$\pm$0.07 & 20.30$\pm$0.05 & 20.03$\pm$0.02 & 0.25 -- 0.60 & 3.8 -- 4.5 &  2.73 -- 1.46& $\cdots$\\
       & 13             & 0.60 & 20.60$\pm$0.07 & 20.20$\pm$0.05 & 19.93$\pm$0.02 & 1.00 -- 1.00 & 5.1 -- 5.5 &  4.06 -- 2.07& $\cdots$\\
\enddata

\tablecomments{Column (1) galaxy name. Column (2) 
identification number of the star clusters/complexes. The $^{\dagger}$ symbol
indicates that Br$\gamma$ emission were measured in the cluster. Column (3) aperture
considered to obtain the photometry. Columns (4), (5) and (6) observed photometry 
in magnitude of each star cluster/complex in the filters J, H and
K$_{p}$, respectively. Columns (7), (8) and (9) extinction E(B-V), age
in units of Myr and the mass in units of M$_{\odot}$ for models I and II,
respectively. Finally Colum (10) give the Br$\gamma$ luminosity.}

\end{deluxetable}

\clearpage

\begin{table*}
\begin{center}
\caption{Structural parameters from the exponential fits to the host galaxies of
 Mrk 36, UM 408 and UM 461. Column (1) galaxy name. 
Columns (2), (3) and (4)  $\mu_{0,(J,H,K_{p})}$,
$\alpha_{0,(J,H,K_{p})}$ parameters and m$_{LSB,(J,H,K_{p})}$ magnitudes. \label{LSB_EXPO}}
\begin{tabular}{cccccc}
\tableline\tableline
Object  &   $\mu_{0,J}$        &  $\mu_{0,H}$        & $\mu_{0,K_{p}}$    \\
        &   (mag arcsec$^{-2}$)&  (mag arcsec$^{-2}$)& (mag arcsec$^{-2}$)\\
        &    $\alpha_{0,J}$    &  $\alpha_{0,H}$     & $\alpha_{0,K_{p}}$ \\
        &      (arcsec)        &  (arcsec)           &    (arcsec)          \\
        &   m$_{LSB,J}$        &  m$_{LSB,H}$        & m$_{LSB,K_{p}}$    \\
        &     (mag)            &      (mag)          &  (mag)             \\
(1)     &  (2)                 &        (3)          &        (4)         \\
\hline
Mrk 36  &     18.64 & 18.25 & 18.38 \\
        &     3.79  &  3.75 &  4.43 \\
        &    14.31 & 13.94  & 13.70 \\
UM 408  &    18.90 & 18.03 & 18.19 \\
        &     1.91 &  1.48 &  1.84 \\
        &    15.89 & 15.58 & 15.27 \\
UM 461  &    19.51 & 19.02 & 19.11  \\
        &     3.69 &  3.37 &  3.84 \\
        &    15.11 & 14.82 & 14.62 \\
\tableline
\end{tabular}
\end{center}
\end{table*}

\clearpage

\end{document}